\documentclass[lettersize,journal]{IEEEtran}
\usepackage{amsmath,amsfonts}
\usepackage{algorithmic}
\usepackage{algorithm}
\usepackage{enumitem}
\usepackage{array}
\usepackage{textcomp}
\usepackage{stfloats}
\usepackage{url}
\usepackage{verbatim}
\usepackage{graphicx}
\usepackage{amssymb}
\usepackage{amsthm} %
\usepackage{comment}
\usepackage{xcolor}
\usepackage{subfiles}
\usepackage[noadjust]{cite}
\usepackage[hidelinks]{hyperref}
\usepackage{mathtools}
\usepackage{setspace}
\usepackage{multirow}
\usepackage{makecell}
\usepackage{hhline}
\usepackage{tablefootnote}
\usepackage[para,online,flushleft]{threeparttable}
\usepackage{makecell, cellspace}
\usepackage{mathrsfs}

\usepackage[caption=false,font=normalsize,labelfont=sf,textfont=sf]{subfig}
\def\NOTES{1}

\if \NOTES 1
 \newcommand{\rsnote}[1]{\textcolor{red}{[RS: #1]}}
 \newcommand{\cknote}[1]{\textcolor{red}{[CK: #1]}}
 \newcommand{\CHANGED}[1]{\textcolor{blue}{#1}}
  \newcommand{\UPDATED}[1]{\textcolor{blue}{#1}} %
 \newcommand{\REVISED}[1]{\textcolor{blue}{#1}}  %
 
\else
 \newcommand{\rsnote}[1]{}
 \newcommand{\cknote}[1]{}
 \newcommand{\CHANGED}[1]{#1}
 \newcommand{\UPDATED}[1]{#1}
 
 \newcommand{\REVISED}[1]{#1}%
\fi

\makeatletter
\AtBeginDocument{\let\CHANGED\@firstofone}
\makeatother

\makeatletter
\AtBeginDocument{\let\UPDATED\@firstofone}
\makeatother

\makeatletter
\AtBeginDocument{\let\REVISED\@firstofone}
\makeatother

\DeclareMathOperator*{\argmax}{arg\,max}

\DeclareMathOperator{\E}{\mathbb{E}}

\definecolor{darkgreen}{rgb}{0.0, 0.5, 0.0}
\theoremstyle{definition} %
\newtheorem{definition}{Definition}%
\newcommand{\norm}[1]{\left\lVert#1\right\rVert} %

\usepackage{soul} %

\begin{document}

\title{Knowledge Transfer for Collaborative Misbehavior Detection in Untrusted Vehicular Environments}

\author{Roshan Sedar, %
Charalampos Kalalas,~\IEEEmembership{Member,~IEEE,}
Paolo Dini, Francisco V\'azquez-Gallego, \\
Jesus Alonso-Zarate,~\IEEEmembership{Senior Member,~IEEE,}
and
Luis Alonso,~\IEEEmembership{Senior Member,~IEEE}
}

\markboth{IEEE Transactions on Vehicular Technology,~Vol.~xx, No.~x, xxx~2024}
{Shell \MakeLowercase{\textit{et al.}}: A Sample Article Using IEEEtran.cls for IEEE Journals}

\maketitle

\begin{abstract}

Vehicular mobility underscores the need for  collaborative misbehavior detection at the vehicular edge.
However, locally trained misbehavior detection models are susceptible to adversarial attacks that aim to deliberately influence learning outcomes.
In this paper, we introduce a deep reinforcement learning-based approach that employs transfer learning for collaborative misbehavior detection among roadside units (RSUs). In the presence of label-flipping and policy induction attacks, we perform selective knowledge transfer from trustworthy source RSUs to foster relevant expertise in misbehavior detection and avoid negative knowledge sharing from adversary-influenced RSUs. The performance of our proposed scheme is demonstrated with evaluations over a diverse set of misbehavior detection scenarios using an open-source dataset. Experimental results show that our approach
significantly reduces the training time at the target RSU and achieves superior detection performance compared to the baseline scheme with tabula rasa learning.
Enhanced robustness and generalizability can also be attained, by effectively detecting previously unseen and partially observable misbehavior attacks.

\end{abstract}
\vspace*{-0.15cm}

\vspace*{-0.2cm}
\section{Introduction}
\vspace*{-0.05cm}
\IEEEPARstart{T}{he} proliferation of connected and automated mobility services, driven by the recent advances in vehicle-to-everything (V2X) technology, leads to a paradigm shift in transport systems. This transformation holds the promise of increased road safety, driving autonomy, and inclusive mobility options.
Inevitably, this %
evolution has given rise to new threat vectors associated with the inherent V2X security vulnerabilities, which adversaries may maliciously exploit to disrupt system operation~\cite{10026338}. %
Among them, misbehavior attacks launched by rogue insiders often become difficult to detect and contain, since malicious nodes may alter their activity intelligently
over time~\cite{van-cITS-2019}.
Although cryptographic techniques are capable of limiting outsiders by offering authentication, integrity, and non-repudiation as a first layer of defense, they may fall short in detecting rogue/dishonest behavior and identifying V2X insiders with malicious intent.
As such, the trustworthiness of exchanged information cannot be guaranteed.

Emerging data-driven approaches fueled 
by artificial intelligence and machine learning (AI/ML) tools provide a fertile ground for addressing misbehavior attacks~\cite{10015746}.
With the availability of vehicular data streams, AI/ML-based schemes can facilitate the analysis of behavioral patterns for V2X entities and determine trustworthiness levels. %
Such capabilities have the potential to overcome the shortcomings of traditional misbehavior countermeasures, offering effective solutions to achieve demanding security requirements.
Centralized detection approaches, albeit leveraging the entire set of available information fused
from multiple geographical locations to reduce the risk of false positives, unintentionally introduce higher computational cost and latency~\cite{8770298}. %
This cogently justifies the use of distributed and decentralized AI/ML-based schemes for collaborative misbehavior detection. By diffusing distributed intelligence across vehicular edge components, such as roadside units (RSUs), resource efficiency gains are sought while detection time can be dramatically reduced.

Despite the envisioned benefits of data-driven misbehavior detection, the vulnerabilities of AI/ML models introduce additional threat vectors, giving rise to finely targeted, stealthy, and scalable \UPDATED{adversarial} attacks. %
Such sophisticated attack types may target both model training (i.e., poisoning
attacks) and test (i.e., evasion attacks) phases~\cite{9019666}, and undermine the efficacy of misbehavior detection.
Collaborative misbehavior detection, relying on decentralized learning models at the vehicular edge, may further extend the attack surface and, thus, exacerbate the impact of adversarial attacks.
Consequently, the pervasive adoption of AI/ML models for misbehavior detection could be hindered if security concerns related to the model vulnerabilities are not addressed.
\subsection{Motivation}
Aiming to address the complex V2X security landscape, several recent works leverage AI/ML-driven techniques for misbehavior detection~\cite{10015746}. Supervised learning schemes may be impractical in V2X scenarios with an expanded attack surface, due to limited access to labeled
training examples and/or dependence on security threshold
values. On the other hand, unforeseen alterations in vehicular mobility, due to
either naturally drifting traffic patterns or unprecedented
malicious activity, introduce challenges (e.g., model overfitting) to misbehavior detection schemes relying on conventional deep learning (DL). In this context, deep reinforcement learning (DRL) methods emerge as a compelling option %
for context-aware misbehavior detection~\cite{10001264}.
By allowing the learner to infer optimal sequential decisions based on rewards/penalties received as a result of previous actions and accumulated experience, the performance of DRL schemes can be dynamically improved in rapidly changing environments.

A number of reputation or trust-based approaches have been proposed in relevant literature to elevate trustworthiness levels in untrusted vehicular environments. Yet, their applicability in collaborative misbehavior detection may be limited, since such methods often assume infrastructure nodes (e.g., RSUs) to be legitimate and non-susceptible to adversarial attacks. Accumulating behavioral patterns of specific nodes over a time period may also be challenging due to ephemeral V2X connectivity links~\cite{hasan-v2x-2020}.
The dependence on predefined trust thresholds may further restrict the suitability of such approaches. %
It is plausible that decentralized learning architectures for collaborative misbehavior detection inadvertently make the underlying models attractive targets for \UPDATED{adversarial attacks}. %
In turn, locally trained misbehavior detection models can be influenced to reach incorrect decisions/predictions or leak confidential information. For instance, in data poisoning attacks~\cite{9681966}, an adversary aims at deliberately modifying the learning algorithm during training via false data injection or manipulation. 
Surprisingly, the detrimental impact of such adversarial manipulations on misbehavior detection performance remains rather unexplored. Dealing with adversarial attacks is often non-trivial and requires
enhanced solutions to foster trust and stimulate confidence in
data-driven misbehavior detection.

\subsection{Contributions}
\CHANGED{Motivated by the research challenges mentioned earlier, our contribution can be summarized in three main aspects}:
\begin{enumerate}
    \item We introduce a novel scheme for collaborative misbehavior detection that builds on geographically distributed RSUs. Each RSU employs a DRL-based model for the detection of malicious traffic stemming from misbehaving vehicles. Leveraging transfer learning principles, the knowledge learned at source RSUs is shared with the target RSU to reuse relevant expertise for misbehavior detection.
    \item In the presence of two data poisoning attacks \CHANGED{with the aim of influencing misbehavior detection outcomes incorrectly,} we perform selective knowledge transfer from trustworthy source RSUs to avoid negative knowledge sharing from adversary-influenced RSUs.
    For this purpose, a novel trust evaluation metric, referred to as semantic relatedness, is used by the target RSU to quantify the trust level of each source RSU for collaborative misbehavior detection.
    \item Considering diverse scenarios of collaborative misbehavior detection and an open-source dataset, we evaluate the learning performance of involved RSUs
    in the presence of adversaries. The detection performance achieved via selective knowledge transfer is also assessed for different misbehavior types. Besides reducing the training time at the target RSU, our scheme is shown to significantly outperform the baseline scheme with tabula rasa learning, demonstrating its high effectiveness. Interestingly, our approach enhances robustness and generalizability by effectively detecting previously unseen and partially observable misbehaviors.
\end{enumerate}%

The remainder of this manuscript is organized as follows. Section~\ref{sec:related_work} provides an overview of the existing work in the related literature. Section~\ref{sec:system_model} outlines the formulation of our proposed misbehavior detection approach, including details about the network and adversarial models considered in the study. Section~\ref{sec:tl_for_drl_mbd} presents the key components of our collaborative misbehavior detection scheme which incorporates transfer learning. Section~\ref{sec:experiments} details the experimental setup and the examined scenarios to validate our approach. Section~\ref{sec:perf_evaluation} evaluates the proposed solution and provides a detailed discussion of the obtained results. Section~\ref{sec:conclusion} summarizes our final remarks and outlines potential directions for future work.         

\section{Related Work}
\label{sec:related_work}
This section summarizes pertinent studies in the literature, with a focus on highlighting open issues and key challenges.

\subsection{Misbehavior Detection}%
\label{ssec:misdet}
Although a multitude of security mechanisms for misbehavior detection are available, existing solutions are \CHANGED{rather} limited in effectiveness due to a \CHANGED{high} %
number of false alarm rates (i.e., false positives and negatives) %
\CHANGED{and lack of robustness} against
sophisticated \UPDATED{adversarial} attacks~\cite{van-cITS-2019,10026338}. Interestingly, %
\CHANGED{such} key open issues underscore the need for advanced data-driven techniques \CHANGED{to achieve} %
effective misbehavior detection.
Data-driven AI/ML approaches 
are a proving ground 
for misbehavior detection in vehicular networks. Recent works in this area, supervised~\cite{so18mlformbd, 9056489, 9662982, prinkle21MLMBD}, unsupervised~\cite{9541028,9474924}, and DRL-based~\cite{10001264, 9838796} \CHANGED{methods}
stand out among others and are \CHANGED{positioned} at the forefront of \CHANGED{the} %
research interest in V2X \CHANGED{security}. %
Next, we summarize existing works on centralized and distributed data-driven AI/ML-based misbehavior detection.

\subsubsection{Centralized Approaches} %
\CHANGED{Several} %
ML-based misbehavior detection \CHANGED{schemes} %
fall into the category of centralized, where models train offline centrally with aggregated training samples collected from multiple vehicles and detect misbehaviors (i.e., downstream task) locally in a decentralized way.  
Centralized ML model training is performed in~\cite{9056489}
using feature vector inputs that originate from the position and movement-based plausibility and consistency checks performed at vehicles. Utilizing trained ML models, all participating nodes perform local detection and share the detected events with \CHANGED{a} central misbehavior authority to make the global decision on misbehaviors.
Results obtained using the open-source VeReMi dataset~\cite{kamel-veremi-20} show that a long short-term memory (LSTM) deep classifier outperforms support vector machines (SVM) and multi-layer perceptron models. 
Another work in~\cite{9656118} utilizes position-related features in VeReMi to train an ensemble supervised learning scheme centrally and detect locally at vehicles or RSUs. %
Numerical results \CHANGED{reveal} %
improved detection performance compared to \CHANGED{benchmark} %
ML-based approaches.   

DL techniques have recently gained widespread attention in cyber-threat detection due to their feasibility and superior performance over traditional ML~\cite{10026338}. %
The work in~\cite{9541028} presents a DL-based unsupervised 
\CHANGED{method}
for identifying misbehaving vehicles. The authors implement various DL architectures by combining different neural network models such as convolutional (CNN), LSTM, and gated recurrent units. In the proposed setup, vehicular messages %
are aggregated and converted into sequences at the RSU and fed to centrally trained DL models on the cloud to detect misbehaving vehicles. %
\CHANGED{Numerical} results based on VeReMi \CHANGED{demonstrate that} %
the proposed DL-based scheme %
\CHANGED{outperforms}
existing ML approaches. In a similar context, an edge-cloud approach is proposed in~\cite{9474924}, which comprises DL engines of CNN and LSTM neural networks for handling misbehaving vehicular traffic. Time-sequence-based and sequence-image-based classification methods are implemented at the edge-cloud to train and detect. Results show that sequence-image-based classification using CNN performs better compared to %
traditional ML-based approaches.

Recent works \CHANGED{in} %
DRL-based misbehavior detection %
~\cite{10001264, 9838796} \CHANGED{stand out from their conventional ML/DL counterparts}
due to their superior detection performance, robustness to noisy training data, and ability to dynamically improve detection \CHANGED{accuracy} %
\CHANGED{in highly volatile V2X} environments. DRL utilizes deep neural networks (DNNs) to approximate \CHANGED{the} state-action value function, which, in turn, the DRL agent effectively learns to map input vehicular traffic to state-action values (Q-values). In~\cite{10001264, 9838796}, messages \CHANGED{transmitted by} %
vehicles are aggregated at RSUs and offload the training of the DRL model onto the cloud server. Misbehavior detection is \CHANGED{then performed} %
at RSUs using the trained model. 
\CHANGED{Performance assessment of the}
proposed DRL\CHANGED{-based} scheme 
\CHANGED{using VeReMi reveals}
\CHANGED{enhanced} %
detection \CHANGED{outcomes} %
compared to \CHANGED{benchmark} %
ML- and DL-based approaches and robustness to noisy training \CHANGED{samples}. %

\subsubsection{Distributed Approaches} %
The \CHANGED{studies} %
that fall into this category execute the offline training of ML models in a distributed or decentralized fashion, either locally in vehicles' on-board units (OBUs) or in RSUs. %
The authors of~\cite{so18mlformbd} exploit the position and movement-based plausibility checks at the vehicle to extract feature vectors and feed them into locally trained supervised ML models such as k-nearest \CHANGED{neighbor} and SVM. %
Results reveal that ML models yield higher accuracy with multiple plausibility checks.
The work in~\cite{prinkle21MLMBD} uses an identical set of plausibility checks used in~\cite{so18mlformbd} to extract relevant features and \CHANGED{feed} %
them into an extended set of supervised ML models. The authors suggest implementing their scheme in OBUs to facilitate local detection in real-time. Numerical results show that \CHANGED{adding} %
plausibility checks improves recall and precision.%

In~\cite{9662982}, the authors train supervised ML models at the RSU by aggregating \CHANGED{basic safety messages (BSMs)} %
from vehicles in \CHANGED{their} communication range. Using an augmented feature set that combines information from successive \CHANGED{BSMs} in VeReMi, they achieve improved detection performance for position falsification attacks compared to existing ML-based approaches. 
Another work in~\cite{8761300} trains a local misbehavior detection system by using supervised ML models to identify false alert transmission and position falsification. By utilizing trained ML models, each vehicle performs local detection and evicts misbehaving vehicles based on aggregated information from all neighboring vehicles. Results show improved detection performance compared to rule- and threshold-based detectors. 

In summary, existing AI/ML-based misbehavior detection models can be either centrally trained in the cloud and tested locally, or both training and detection are realized in vehicles/RSUs.
Yet, these models 
are susceptible to adversarial attacks, e.g., data poisoning \cite{chakraborty2018adversarial}. As such, an attacker may poison the centralized model training pipeline and influence downstream tasks at vehicles or RSUs to misclassify, which could inflict a single point of failure. Similarly, locally trained misbehavior detection models may suffer from data poisoning and adversarial manipulations \CHANGED{owing to the extended attack surface}. Hence, albeit offering enhanced detection performance, such models may not be robust enough against adversarial attacks. Moreover, they may have limited capability in detecting unseen 
\CHANGED{(i.e., non-anticipated)} misbehavior attacks. \REVISED{On the contrary, we propose distributed collaborative learning to enhance misbehavior detection performance, making it robust against adversarial attacks and capable of generalizing to detect unseen attacks.}

\subsection{Trust Evaluation} %
\label{ssec:trust}
Data-driven misbehavior detection models analyze the transmitted messages \CHANGED{either} individually at \CHANGED{a} %
local level or collaboratively at a global level. They focus on verifying the semantic correctness of the received information regardless of the sender and do not rely on the honest majority of senders. Entity-centric models, on the other hand, focus on monitoring the past behavior of nodes over time to assess their trustworthiness~\cite{van-cITS-2019,hasan-v2x-2020}. Such models primarily constitute trust-based detection techniques where each entity maintains a local table to record the reputation or trust indicators of its neighboring vehicles, and they often assume the majority of senders \CHANGED{to be} honest. %

In~\cite{li-art-2016}, the authors evaluate the trustworthiness of vehicular data and nodes using the Dempster-Shafer belief framework. A vehicle's trust is evaluated on both functional and recommendation trust. %
\CHANGED{Functional} trust directly evaluates the trustworthiness of a vehicle, and recommendation trust evaluates the trustworthiness of neighboring vehicles. Results \CHANGED{reveal} %
superior performance compared to a conventional weighted-voting method \CHANGED{for} %
trust management. Another similar work in~\cite{kerrache17tfdd} presents a trust-based mechanism against DoS/DDoS attacks, with vehicles assessing neighboring senders via direct and indirect trust values.
In general, these mechanisms can be integrated with data-driven AI/ML-based methods to enhance trustworthiness in misbehavior detection.   

The work in~\cite{9003228} proposes a context-aware data-driven trust management scheme to ascertain the integrity of information received by vehicles. The authors employ information entropy theory to calculate the trustworthiness of the content of messages and incorporate an \CHANGED{RL}
model based on $Q$-learning to continuously adapt the trust evaluation strategy to accommodate various driving scenarios. %
Numerical results indicate the ability to withstand bogus information, \CHANGED{outperforming} %
entity-centric trust schemes. 
Another work in~\cite{9428520} presents a DRL-based dynamic reputation update mechanism for vehicular 
networks. The authors apply the Dempster-Shafer theory to combine misbehavior detection reports sent by vehicles centrally at \CHANGED{a} global level and train a DRL model to update the reputation policy. The DRL model dynamically determines the \CHANGED{optimal} reputation update value to stimulate vehicles to send true feedback. Simulation results  \CHANGED{exhibit improved} %
performance compared to entity-centric trust schemes.   

In summary, most reputation or trust-based methods in related literature are entity-centric and often rely on infrastructure assistance (e.g., RSUs), assuming the presence of an honest majority of participants. Furthermore, these methods frequently resort to predefined conditions, ranging from task-specific rule sets to statically configured threshold values, and \CHANGED{assume} %
infrastructure nodes \CHANGED{to be} %
benign and immune to adversarial attacks. In addition, monitoring \CHANGED{nodes} %
over an extended period to evaluate trustworthiness \CHANGED{often becomes} %
challenging due to the transient connections and mobility patterns of vehicles. Overall, trust management models are not \CHANGED{sufficiently robust against} adversarial attacks. \REVISED{However, our proposed approach in this work does not rely on predefined conditions, such as an honest majority of participants or statically configured threshold values, to determine trustworthiness.}

\begin{figure*}[!t]
\centering
\includegraphics[width=0.8\textwidth]{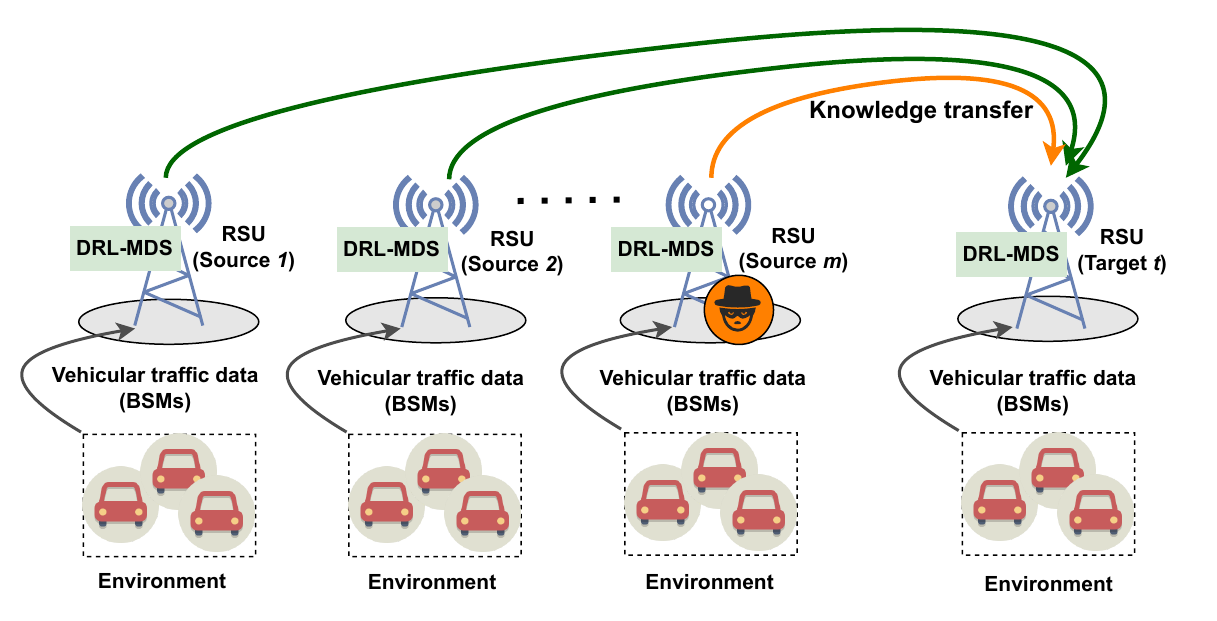}
\vspace*{-0.6cm}
\caption{Considered network model for collaborative misbehavior detection. %
Distributed knowledge transfer between source RSUs $\{s_{i}\}^m_{i=1}$ and the target RSU $t$ is depicted with %
\CHANGED{green} arrows %
for positive knowledge and \CHANGED{with} orange for negative knowledge. 
\CHANGED{The presence of a malicious adversary implies}
that \CHANGED{the} training of the misbehavior detection system (DRL-MDS) at \CHANGED{certain} source RSUs (e.g., RSU $m$) is under adversarial influence. %
}
\label{fig:network_model}
\vspace*{-0.42cm}
\end{figure*}

\subsection{Adversarial Attacks}
\CHANGED{With the increasing penetration of data-driven methods into V2X systems},
AI/ML models \CHANGED{inevitably become} %
\CHANGED{prone} %
to adversarial \CHANGED{manipulations}%
~\cite{chakraborty2018adversarial}. In an adversarial ML setting, attackers inject adversary data intending to misguide AI/ML models to incorrect decisions. In this context, the prevalence of adversarial attacks hinders the robustness of AI/ML models, \CHANGED{while} %
dealing with \CHANGED{such} %
attacks is often non-trivial. Inherently, previously described misbehavior detection (Sec.~\ref{ssec:misdet}) and trust (Sec.~\ref{ssec:trust}) models are vulnerable to adversarial ML. %

Poisoning and evasion attacks can be identified as the main \CHANGED{adversarial} attack scenarios~\cite{chakraborty2018adversarial}. The contamination of training data, %
known as poisoning, %
occurs during the training \CHANGED{phase} %
of the AI/ML model. In this type of attack, the attacker \CHANGED{attempts} %
to inject an imperceptible perturbation \CHANGED{into} the input data, which corrupts \CHANGED{training} %
and misguides the AI/ML model~\cite{goodfellow2014explaining}. %
Similarly, label-flipping can also be used as a poisoning attack, \CHANGED{yielding a} simple and effective way to contaminate the training data~\cite{lin2021ml}. In this attack, the adversary selects a subset of training samples and flips their labels. In contrast, evasion attacks take place at the testing \CHANGED{phase}, %
where the adversary has no influence on the target model during training, but rather tricks the model into incorrect outputs. %

The work in~\cite{9032989} demonstrates the vulnerability of ML-based misbehavior detection models against adversarial attacks. The authors inject carefully crafted adversarial examples using ML and DL techniques and compromise the detection model to misclassify attack samples. In this approach, ML and DL models are first trained on the genuine behavior dataset, followed by generating adversarial examples. The evaluation using VeReMi dataset \CHANGED{reveals} %
high false positive and false negative values. Another work in~\cite{8685687} %
\CHANGED{introduces} a data poisoning attack in DL models, %
which is equipped with attacker-chosen patterns to force misclassification. %
Such attacks can take place when the training is outsourced onto the cloud or when using pre-trained models in transfer learning. Similarly, DRL-based models are also \CHANGED{susceptible} %
to adversarial attacks. %
In~\cite{behzadan2017vulnerability}, the authors present a policy induction attack by injecting crafted adversarial examples into the agent's state space. %
Results show that the agent's average reward per epoch remains at low levels in the presence of adversarial examples.

\REVISED{\textbf{Summary:}} Considering the \CHANGED{open research gaps highlighted throughout the previous} state-of-the-art overview, we \CHANGED{hereby} focus on collaborative misbehavior detection by transferring the knowledge learned between distributed entities in untrusted vehicular environments.   
Our proposed DRL-based scheme can select trustworthy RSUs to collaborate and transfer knowledge withstanding adversarial attacks. Consequently, target or receiving entities can learn efficiently from transferred knowledge to discover unseen and partially observable misbehaviors.

\section{System Model}
\label{sec:system_model}
This section introduces the vehicular network model, misbehavior detection model, and adversarial model considered in this work. \CHANGED{We provide the details in the following.}
\subsection{Network Model}
\CHANGED{Vehicular networks typically comprise} a
large number of geographically distributed RSUs. %
RSUs are
stationary entities interconnected with each other and the Internet. The usability of RSUs is \CHANGED{multifaceted} %
as they offer various services such as Internet access, security solutions, and real-time traffic data distribution~\cite{10026338}. Normally, RSUs %
have superior computational \CHANGED{capabilities} %
\CHANGED{compared to} %
in-vehicle resources, which allows vehicles to offload %
computation-\CHANGED{intensive} %
tasks to RSUs. 
Fig.~\ref{fig:network_model} \CHANGED{illustrates} %
the distributed vehicular network model \CHANGED{considered} in this work. %
Each RSU \CHANGED{experiences} %
its own vehicular environment and receives traffic %
within its coverage. %
\UPDATED{The RSUs are interconnected via wired connections to provide reliable RSU-to-RSU communication.} We assume authentication and authorization \CHANGED{procedures have already been} performed before V2X communication \CHANGED{takes place} %
among entities~\cite{scms-brecht-2018}.
The involved vehicles (i.e., authenticated and authorized) periodically broadcast BSMs, which \CHANGED{are} %
\CHANGED{received by} 
RSUs in \CHANGED{their} %
connectivity range. BSMs include standard\CHANGED{-related} parameters such as position, speed, acceleration, heading angle, and other relevant \CHANGED{vehicular} information~\cite{kamel-veremi-20}.     

In our proposed \CHANGED{setup}, %
a distributed collaborative misbehavior detection system is considered where we leverage knowledge transfer in the context of transfer learning. As shown in Fig.~\ref{fig:network_model}, each RSU is equipped with a DRL-based misbehavior detection system (DRL-MDS) and detects misbehaving vehicles. 
Misbehavior detection is %
performed at the RSU \CHANGED{level,} since the vehicle may not have complete information in its communication range
due to ephemeral
connectivity and/or limited computational resources. 
A high-level illustration of the DRL-MDS workflow is %
provided in Fig.~\ref{fig:drl_mds_rsu},  
\CHANGED{and it is} \CHANGED{further elaborated} %
in the following section (Sec.~\ref{ssec:MisDet_model}). The knowledge learned at source RSUs is transferred to the target RSU to reuse existing knowledge during its learning process. The target RSU may not necessarily need to be a neighbor, but it could also be a distant RSU, which shall be reusing the available knowledge of sources to detect future misbehaviors. 
\CHANGED{It is noteworthy that} RSU-based security solutions in related literature tend to assume that RSUs are trusted entities and non-vulnerable to attacks~\cite{hasan-v2x-2020}. However, as pointed out in Sec.~\ref{sec:related_work}, stationary entities such as RSUs can easily be targeted by adversarial attackers. In this work, we assume that there may exist RSUs that are subject to adversarial attacks.         
\subsection{Misbehavior Detection Model} 
\label{ssec:MisDet_model}

\begin{figure*}[!t]
\centering
\includegraphics[width=0.65\textwidth]{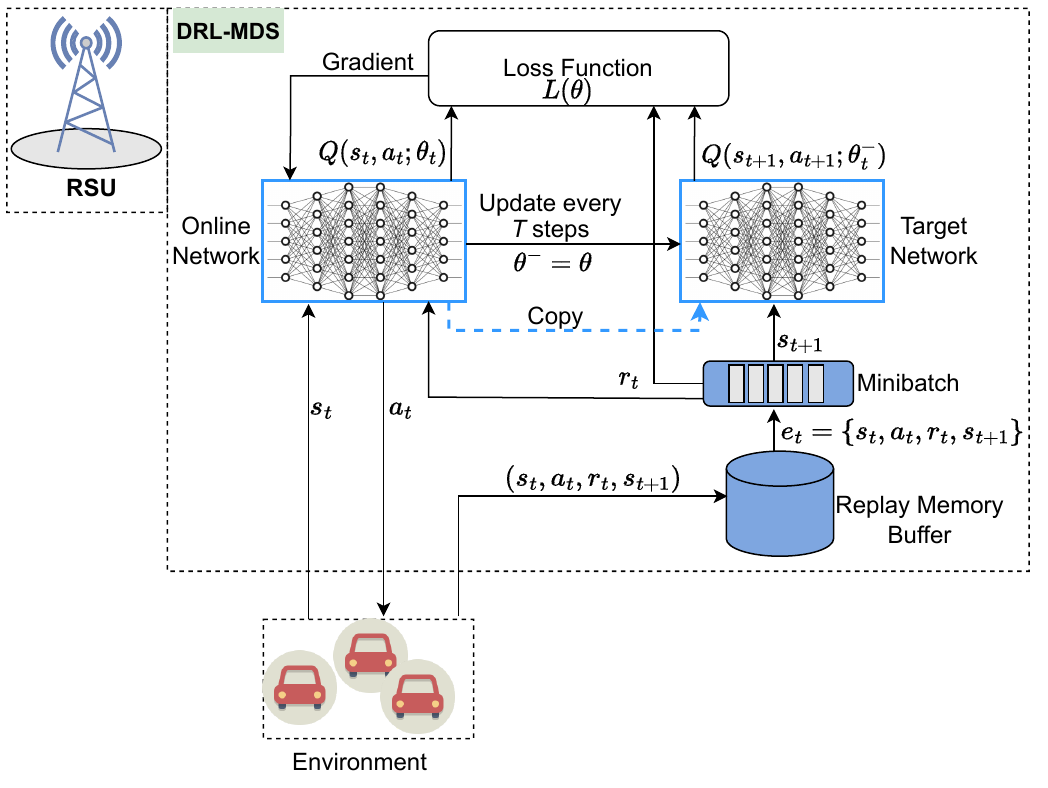}
\caption{High-level illustration of the DRL-MDS workflow in an RSU. \REVISED{In every agent-environment interaction, the tuple ($e_{t} = <s_{t}, a_{t}, r_{t}, s_{t+1}>$) is stored in the replay memory buffer as the agent’s instantaneous experience, and a random minibatch of past experiences is sampled to regularize the training process. The online and target networks comprise the same DNN structure, with the target network being updated periodically (every $T$ steps, copy weights of $\theta$ to $\theta^-$) using the weights from the online network to keep it in sync with the online network}.}
\label{fig:drl_mds_rsu}
\end{figure*}

In our approach, each RSU is equipped with an MDS as shown in Fig.~\ref{fig:drl_mds_rsu}. The DRL-based model, introduced in our previous works~\cite{9838796, 10001264}, is used for detecting misbehaving vehicles. Here, we present the formulation of the tabula rasa DRL model used for misbehavior detection. A tabula rasa model aims to learn efficiently from scratch without any external/previous knowledge. The details of transfer learning incorporation into the DRL model are provided in Sec.~\ref{sec:tl_for_drl_mbd}.

\subsubsection{DRL Model} 
\label{sssec:RLmodel}
\UPDATED{The vehicular environment considered in this study follows a Markov decision process (MDP) framework to facilitate the detection of misbehaviors through sequential decision-making. Consistent with the MDP formulation, the action of misbehavior detection changes the environment based on the decision of either genuine or malicious behavior at time-step $t$. Subsequently, the decision at time-step $t+1$ is influenced by the altered environment from the previous time-step $t$.} 
The aggregated vehicular traffic at each RSU consists of a time-series repository of received BSMs with intrinsic spatiotemporal interdependencies.  In this work, we consider a DRL-based misbehavior detector deployed at the edge RSU. As depicted in Fig.~\ref{fig:drl_mds_rsu}, the detector (agent) interacts with the vehicular environment to learn the optimal detection policy $\pi^*$. %
\UPDATED{During training, the $\epsilon$-greedy method is leveraged to strike a balance between exploration and exploitation in the agent's strategy. Next, we describe the components relevant to the DRL model.}

\textit{i) Agent:} The agent receives the vehicular traffic data as a time series and prior related decisions as inputs (i.e., state $\REVISED{\mathbf{s_{t}}}$), and generates the new decision made (i.e., action $a_{t}$) as output, as shown in Fig.~\ref{fig:drl_mds_rsu}. The agent's DQN~\cite{mnih2013playing} comprises an LSTM layer and a fully connected neural network with linear activation to generate $Q$-values as choices for the action $a_{t}$. At each time-step $t$, the agent's actions are selected by the policy $\pi$. The agent's experience, i.e., $ e_{t} = <\REVISED{\mathbf{s_{t}}},a_{t},r_{t},\REVISED{\mathbf{s_{t+1}}}> $, stores all the behaviors of the misbehavior detector. By exploiting experience, the detector is progressively improved to obtain a better estimation of the $Q(\REVISED{\mathbf{s}},a)$ function. The objective is to maximize the expected sum of future discounted rewards by learning the $\pi^*$. The discounted reward return is expressed as 
\setlength{\abovedisplayskip}{5pt}
\setlength{\belowdisplayskip}{5pt}
\begin{align}
R_{t} = \sideset{}{}\sum_{k=t}^{T} \gamma^{k-t}r_{k}.
\end{align}%
\normalsize
$Q$-learning model updates are performed with learning rate $\alpha$ and discount factor $\gamma$ as
\begin{align}
\begin{split}
Q(\REVISED{\mathbf{s_{t}}}, a_{t}) &\leftarrow Q(\REVISED{\mathbf{s_{t}}}, a_{t})\\
 &+\alpha(r_{t} + \gamma\max_{a_{t+1}} Q(\REVISED{\mathbf{s_{t+1}}}, a_{t+1}) - Q(\REVISED{\mathbf{s_{t}}}, a_{t})).
\end{split}
\end{align}%

\textit{ii) States:} The state space comprises the sequence of previous actions denoted by $s_{action} = <a_{t-1},a_{t},...,a_{t+n-1}>$, and the current BSM information denoted by $s_{time} = <X_{t},X_{t+1},...,X_{t+n}>$. $\REVISED{\mathbf{X_{t}}} \in \mathbb{R}^d$ is a $d$-dimensional feature vector at time-step $t$, including information on $d$ different features. According to the state design, the next action taken by the agent depends on the previous actions and the current BSM information. This design enables the agent to capture temporal dependencies and make more informed decisions. 

\textit{iii) Actions:} The action space is defined as $\mathcal{A}$ = \{0,1\}, where $1$ indicates the detection of \UPDATED{a misbehavior} and $0$ represents the genuine behavior. The deterministic detection policy $\pi$ can be expressed as a mapping, i.e., $\pi:\REVISED{\mathbf{s_{t}}} \in \mathcal{S} \longmapsto a_{t} \in \mathcal{A}$,  from states to actions, where $\pi(s)$ prescribes the action that the agent takes at state $\REVISED{\mathbf{s}}$.  
In a given state $\REVISED{\mathbf{s_{t}}}$, the agent selects the action based on the optimal detection policy given by
\begin{align}
        \pi^* = \argmax\limits_{a \in \mathcal{A}} Q^*(\REVISED{\mathbf{s}},a).
        \label{eq:eps_greedy}
\end{align}

\textit{iv) Rewards:} %
The reward function $R(\REVISED{\mathbf{s}},a)$ is defined based on the confusion matrix typically used in ML classification problems. A numerical value for $r_{t}$ is assigned based on the ground truth information of BSMs. A positive reward is given upon correct detection of a \UPDATED{misbehavior}, i.e., true positive (TP), or a normal state, i.e., true negative (TN). A negative reward is given when a normal state is incorrectly identified as a \UPDATED{misbehavior}, i.e., false positive (FP), or a \UPDATED{misbehavior} as a normal state, i.e., false negative (FN). The agent is penalized more for FN actions than for FPs, as the correct identification of misbehavior is necessary to avoid hazardous situations. Accordingly, we define the immediate reward $r_{t} \in R$ of the agent as
\setlength{\abovedisplayskip}{0pt}
\setlength{\belowdisplayskip}{5pt}
\begin{align}
r(\REVISED{\mathbf{s_{t}}},a_{t}) = \begin{cases}
                a, & \text{if}~a_{t}~\text{is a TP,}\\
                b, & \text{if}~a_{t}~\text{is a TN,}\\
                -c, & \text{if}~a_{t}~\text{is an FP,}\\
                -d, & \text{if}~a_{t}~\text{is an FN,}
            \end{cases}
\label{eq:rew_func}
\end{align}
where $a, b, c, d > 0$,  with $a > b$ and $d > c$. 

\subsection{Adversarial Model}
\CHANGED{In this work,} 
we assume the presence of adversaries attempting to contaminate the training data and realize poisoning attacks on the distributed and collaborative DRL-MDS models. In particular,
we consider two data poisoning attacks pertinent to adversarial ML: $i)$ label-flipping and $ii)$ policy induction \CHANGED{attacks}. 
In the case of label-flipping, it is assumed that the adversaries are rogue insiders who contribute to the training data or have access \CHANGED{to} %
the training data itself. In policy induction, we consider an exogenous attacker who can modify the state space before it is observed by the DRL agent. In both cases, attackers aim to maliciously influence the learning model and, subsequently, force incorrect outcomes in downstream misbehavior detection tasks. %

In \CHANGED{a} label-flipping attack, we consider an %
attacker who targets the malicious class to flip the labels of certain
source training data instances at an RSU. In this scenario, the adversarial attacker randomly selects a set of misbehaving vehicles and flips the labels of their data instances into genuine ones, resulting in a targeted random label-flipping attack (Definition~\ref{def:label-flip}). The attacker aims to misclassify selected training samples from malicious class $1$ to genuine class $0$. Following Definition~\ref{def:label-flip}, we denote $D_{S_{i}}$ as the training data of source RSU $S_{i}$, i.e., $D_{S_{i}} = \{(x_{S_{i1}}, y_{S_{i1}}),..., (x_{S_{ik}}, y_{S_{ik}}),..., (x_{S_{in}}, y_{S_{in}})\}$, with $x_{S_{ik}} \in X_{S_{i}}$ representing the $k$-th data instance of $D_{S_{i}}$ while $y_{S_{ik}} \in Y_{S_{i}}$ represents the corresponding label of $x_{S_{ik}}$. %
\begin{definition}[\textbf{Label-flipping attack}]
\label{def:label-flip}
Given training samples $\{x_{S_{ik}}, y_{S_{ik}}\}^n_{k=1}$ of $D_{S_{i}}$, belonging to source RSU $S_{i}$ with $x_{S_{ik}} \in X_{S_{i}}$ and $y_{S_{ik}} \in Y_{S_{i}}= \{0, 1\}$,
a poisoning attack is defined as
a targeted random label-flipping attack when the attacker randomly selects \CHANGED{a} fraction $\zeta \in (0, 1]$ of misbehaving vehicles and \CHANGED{flips} the label of the corresponding training samples \CHANGED{from $1$ to $0$}.  
\end{definition}
In \CHANGED{a} policy induction attack, we \CHANGED{properly} adapt the adversarial attack originally presented in~\cite{behzadan2017vulnerability} in a DRL context. %
The malicious intent here is to force the target DRL agent
to learn a policy selected by the adversary.
We assume an exogenous attacker with minimal \textit{a priori} information (e.g., input type and format) of the target DQN,
and with knowledge of its reward function and the update frequency of the target %
network. The attacker can directly manipulate the target DQN's environment configuration, albeit with 
no control over the target network parameters, reward function, or optimization mechanism. According to 
Definition~\ref{def:adv_policy}, the attacker creates replica DQNs (i.e., $Q'$, $\hat{Q}'$) of the target's DQN (i.e., $Q$, $\hat{Q}$) and initializes them with random parameters. Since the attacker has no knowledge of the target's DQN architecture and its parameters at every time step, a black-box technique is followed to exploit the transferability of adversarial examples. This is achieved by crafting state perturbations using replicas of the target's DQN such as those introduced in~\cite{papernot2017practical}. Moreover, the Fast Gradient Sign Method (FGSM)\footnote{%
The FGSM method utilizes the neural network gradients' sign to create a perturbed adversarial input that maximizes the loss~\cite{goodfellow2014explaining}. For a given input $x$, the adversarial input $x^\prime$ generated by the FGSM can be summarized as $x^\prime = x + \epsilon * \text{sign}(\nabla_{x}J(\REVISED{\mathbf{\theta}},x,y))$, where $y$ is the label of input $x$, $\epsilon$ denotes a small multiplier, $\REVISED{\mathbf{\theta}}$ are the model parameters, and $J$ represents the loss.} algorithm is used to craft adversarial examples at every training time step. 
\begin{definition}[\textbf{Policy induction attack}]%
\label{def:adv_policy}
The attacker induces an arbitrary adversarial policy $\pi_{adv}$ on the target DQN by injecting adversarial examples into training data. Specifically, given state observation $\REVISED{\mathbf{s_{t+1}}}$, the attacker crafts a perturbation vector ($\REVISED{\mathbf{\hat{\delta}_{t+1}}}$) using FGSM at every training time-step and injects it into the target DQN's state space, as
\setlength{\abovedisplayskip}{5pt}
\setlength{\belowdisplayskip}{5pt}
    \begin{align}
        a'_{adv} &\leftarrow \pi_{adv}(\REVISED{\mathbf{s_{t+1}}}),\\
        \REVISED{\mathbf{\hat{\delta}_{t+1}}} & \leftarrow Craft(\hat{Q}',a'_{adv},\REVISED{\mathbf{s_{t+1}}}),\\
        \vspace{-0.25cm}
         \REVISED{\mathbf{s'_{t+1}}} & \leftarrow \REVISED{\mathbf{s_{t+1}}} + \REVISED{\mathbf{\hat{\delta}_{t+1}}}.
         \vspace{-0.25cm}
     \end{align}%
\noindent Crafting adversarial inputs $\REVISED{\mathbf{s'_{t}}}$ and $\REVISED{\mathbf{s'_{t+1}}}$ requires minimizing the loss function in (\ref{eq:loss}) to force the target DQN to optimize towards action $a'$, given state $s_t$, as
    \begin{align}
        min_{\REVISED{\mathbf{\theta'}}} (y_{t} - Q'(\REVISED{\mathbf{s_{t}}},a_{t};\REVISED{\mathbf{\theta'}}))^2,
        \label{eq:loss}
    \end{align}
\noindent where $y_{t} = r_{t} + \gamma max_{a'} \hat{Q}' (\REVISED{\mathbf{s'_{t+1}}},a';\REVISED{\mathbf{\theta'_{-}}})$ and $Q^{'}$, $\hat{Q^{'}}$ are the replica DQNs of the target DQN.
\end{definition}

\section{Transfer Learning-Based DRL Framework} %
\label{sec:tl_for_drl_mbd}

\begin{figure*}[!t]
\centering
\includegraphics[width=0.95\textwidth]{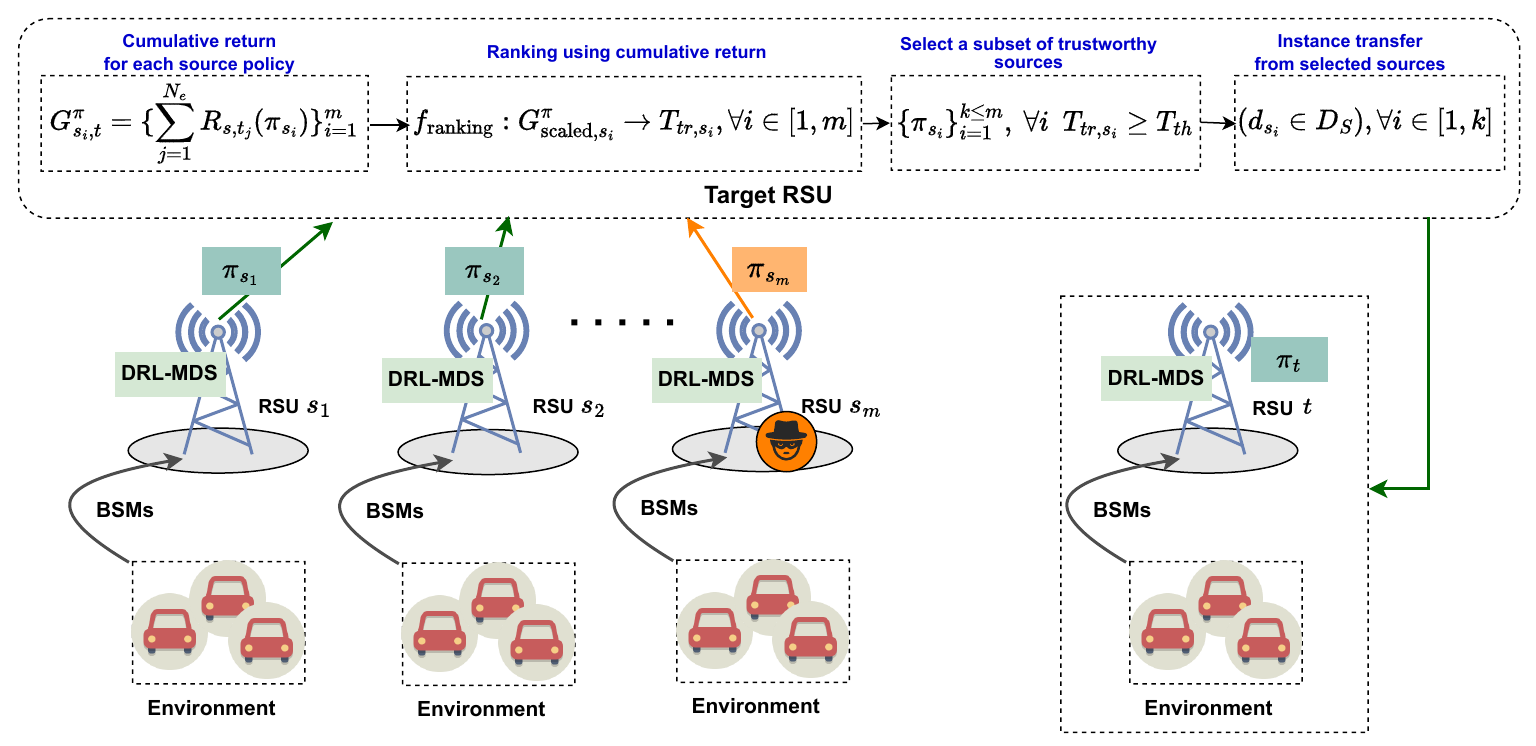}
\caption{Visualization workflow for selecting trustworthy source RSUs. Source RSUs with policies $\{\pi_{s_{i}}\}^m_{i=1}$ are ranked ($\mathscr{f}_{\text{ranking}}$) based on normalized scale cumulative return values $\{G^\pi_{\text{scaled}, s}\}^m_{i=1} \in [0,1]$, which are extracted from the target RSU $t$'s environment. Instances $\{(s^{(t)}_{i},a^{(t)}_{i},r^{(t)}_{i},s^{(t+1)}_{i}) \coloneqq d_{s_{i}}\}^k_{i=1}$ are transferred to the target RSU $t$ from $k$ subset of trustworthy source RSUs with policies $\{\pi_{s_{i}}\}^{k}_{i=1}$, where $k \leq m$ and $D_{S}$ denotes samples of source RSUs. Green arrows denote positive transfer from trusted sources and the orange arrow represents a potential negative transfer from a malicious source.} %
\label{fig:selective_transfer_scheme}
\end{figure*}

The distributed RSU deployments in vehicular %
systems \UPDATED{and} the varying \CHANGED{spatiotemporal}
behavior of %
traffic traces, 
render the training of a DRL-MDS agent difficult, especially in
detecting unseen and partially observable \UPDATED{misbehavior} attacks. 
Challenges in such complex setups include slow convergence of the learning process, overfitting, and/or sub-optimal solutions due to poor exploration~\cite{yuan2021multimodal}.
Hence, we advocate %
the adoption of a transfer learning-based DRL approach to achieve distributed collaborative misbehavior detection in adversarial vehicular environments. We rely on selectively transferring knowledge from trustworthy source RSUs to avoid negative knowledge sharing from adversary-influenced RSUs, as shown in Fig.~\ref{fig:selective_transfer_scheme}. %

\subsection{Transfer Learning for DRL-based Misbehavior Detection}
\label{ssec:tl_drl_definition}
TL leverages valuable knowledge acquired in one task and past experience to improve learning performance on other tasks (similar/dissimilar). The knowledge learned and previous experiences obtained from learning some \textit{source} tasks, can be re-used to enhance the learning of some \textit{target} tasks. In our work, the task represents the misbehavior detection performed in each RSU. %
Most DRL-based works in related literature (Sec.~\ref{sec:related_work}) have developed agents that can efficiently learn tabula rasa without any previously learned knowledge. Yet, tabula rasa DRL techniques require a long learning period to be effective, which can be inefficient in computationally intensive and mission-critical \CHANGED{operations}~\cite{10172347}. 
For instance, an agent at the start of learning may tend to spend significant time exploring the environment before finding an optimal policy. Thus, TL can be leveraged to accelerate the learning.
In this context, source and target RSUs associated with TL can be represented as MDP\footnote{Henceforth, MDP and environment are used interchangeably.} agents. Accordingly, TL between RSUs can be defined as follows.

\begin{definition}[\textbf{Transfer learning between RSUs}]
\label{def:TL_DRL}
Given a set of $m$ source RSUs with MDPs $\mathbf{M}_{s} = \{\bigcup\limits_{i=1}^{m} \mathcal{M}_{i,s}\rvert\mathcal{M}_{i,s} \in \mathbf{M}_{s}\}$ and a target RSU with MDP $\mathcal{M}_{t}$, the goal of TL is to learn an optimal policy $\pi^*_{t}$ for the target RSU as 
\setlength{\abovedisplayskip}{5pt}
\setlength{\belowdisplayskip}{5pt}
        \begin{align}
             \pi^*_{t} = \argmax\limits_{\pi_{t}} \E_{\REVISED{\mathbf{s}} \sim \REVISED{\mathbf{\mathcal{S}_{t}}},a \sim \pi_{t}} [Q^{\pi_{t}}_{\mathcal{M}_{t}}(\REVISED{\mathbf{s}},a)],
             \label{eq:opt-policy}
    \end{align}
\noindent 
by leveraging external information $\mathcal{I}_{s}$ from $\mathbf{M}_{s}$ along with internal information $\mathcal{I}_{t}$ from $\mathcal{M}_{t}$.
In (\ref{eq:opt-policy}), %
$\pi_{t} = \phi(\mathcal{I}_{s} \sim \mathbf{M}_{s}, \mathcal{I}_{t} \sim \mathcal{M}_{t}):\REVISED{\mathbf{s_{t}}} \in \mathcal{S} \rightarrow a_{t} \in \mathcal{A}$ is a policy that maps states to actions for $\mathcal{M}_{t}$, which is learned from both $\mathcal{I}_{t}$ and $\mathcal{I}_{s}$. The policy $\pi_t$ is estimated using a DNN.
\end{definition}
In multi-source\footnote{In a single-source scenario with
$\lvert \mathbf{M}_{s} \rvert = 1$,  tabula rasa learning occurs without any external transfer when $\mathcal{I}_{s}=\emptyset$.} RSU deployments, according to Definition~\ref{def:TL_DRL}, the aim is to enhance the learning of misbehavior detection at the target RSU by leveraging knowledge acquired from a set of source RSUs. 
Assume 
$m$ source RSUs with MDPs $\mathbf{M}_{s} = \{(\mathcal{S}_{i},\mathcal{A}_{i},\mathcal{P}_{i},\mathcal{R}_{i},\gamma_{i})\}^{m}_{i=1}$, and a detection policy $\pi_{s_{i}}$ for each source RSU $i$. %
Then, the target RSU with MDP $\mathcal{M}_{t} = (\REVISED{\mathbf{\mathcal{S}_{t}}},\mathcal{A}_{t},\REVISED{\mathbf{\mathcal{P}_{t}}},\mathcal{R}_{t},\gamma_{t})$ %
aims to enhance the learning of $\pi^*_{t}$ by leveraging %
$\mathcal{I}_{s}\sim \mathbf{M}_{s}$. The external $\mathcal{I}_{s}$ %
represents transferred knowledge on misbehavior detection from source RSUs. %
\subsection{Knowledge Transfer}
\label{ssec:tl_drl_mds}
Previous works (e.g., \cite{tao2021repaint,gupta2016learning,lin2017collaborative}) on the integration of TL with collaborative DRL have demonstrated effectiveness in scenarios with some degree of similarity in agents' experiences. %
Based on the definition of transferred knowledge, TL techniques in DRL can be realized as policy\footnote{In policy transfer, a set of policies $\{\pi_{s_{i}}\}^{m}_{i=1}$ from source MDPs are transferred to the target MDP, and then the target policy $\pi_{t}$ is learned by utilizing knowledge from them.}, representation\footnote{In representation transfer, the algorithm learns a feature representation of the task, such as a value function $V^{\pi}(\REVISED{\mathbf{s}})$ or the $Q^{\pi}(\REVISED{\mathbf{s}},a)$ function, and the knowledge learned can be either directly used in the target or indirectly using a task-invariant feature space.}, and instance transfer~\cite{10172347,garcia2022taxonomy}.
We hereby employ an instance TL technique to transfer misbehavior detection knowledge from a set of source RSUs to %
a target RSU.  %
The rationale lies in transferring those
source samples that can enhance the detection performance of the target RSU.
Our method
selects only related (i.e., good/useful) samples collected from the environments of source RSUs as agents' experiences. The selection of such samples prevents negative knowledge transfer that may originate from adversarial source RSUs. In collaborative DRL-MDS,
positive transfer occurs when the transferred knowledge from source RSUs improves 
the target RSU's performance. 
In negative transfer, the target's performance degrades compared to tabula rasa learning. %

To realize collaborative DRL-MDS, our proposed methodology employs TL with a new selective knowledge transfer scheme to prevent negative transfers from untrusted source RSUs, as in Fig.~\ref{fig:selective_transfer_scheme}. Under positive transfer, the target RSU is expected to achieve higher cumulative return as well as reach the asymptotic performance earlier than in tabula rasa learning. Hence, measuring the degree of similarity in misbehavior detection performed at the source and target RSUs
becomes crucial to avoid negative transfer overhead in TL.

\subsubsection{Trust Evaluation} In Fig.~\ref{fig:selective_transfer_scheme}, we provide a visualization workflow for trustworthy source RSUs' selection for transfer. 
We first establish an inter-agent semantic similarity metric between the source and target to ensure a positive transfer. 
A source RSU $s_i$ with policy $\pi_{s_{i}}$ is considered to have
high \textit{semantic relatedness} with the target RSU $t$, if $\pi_{s_{i}}$ can generate the maximum cumulative return from the target RSU's environment in a limited number of training episodes. 
Inspired by works in~\cite{tao2021repaint,lotfi2022semantic}, the semantic relatedness %
between the source and target RSUs is defined %
as the gain of cumulative return for %
a policy $\pi_{s_{i}}$ under the target reward function, 
\setlength{\abovedisplayskip}{3pt}
\setlength{\belowdisplayskip}{3pt}
\begin{align}
     G^\pi_{s_{i}} = \sum^{N_{e}}_{j=1} R_{t_{j}}(\pi_{s_{i}}), \quad \forall i \in [1,m],
     \label{eq:semantic_similarity}
\end{align}
\noindent where $N_{e}$ denotes the number of training episodes and $R_{t}(.)$ denotes the target reward function. 

By utilizing semantic relatedness, the target RSU can effectively select a subset of source RSUs that are trustworthy to contribute towards enhanced misbehavior detection. %
A higher cumulative return in $N_{e}$ training episodes ascertains a positive transfer of detection knowledge and implies that a source RSU is more reliable for collaboration.
The procedure of selecting trustworthy source RSUs for knowledge transfer is summarized in Algorithm~\ref{source_ranking}. Without loss of generality, we normalize cumulative return $G^\pi_{s_{i}}$ in a way that the scaled cumulative return $G^\pi_{\text{scaled}, s_{i}}$ resides in the range of $[0,1]$.

\setlength{\textfloatsep}{0pt}%
\begin{algorithm}[!ht]
\footnotesize
\caption{Selection of trustworthy source RSUs for transfer}
\label{source_ranking}
\begin{algorithmic}[1]
    \REQUIRE Load $\pi_{s_{1}}, \pi_{s_{2}},...,\pi_{s_{m}}$ , where $i \in [1,m]$ %
    \ENSURE Trusted policies $\pi_{s_{1}}, \pi_{s_{2}},...,\pi_{s_{k}}$, where $k \leq m$
    \FOR{iteration $i$ = 1 to $m$}
       \vspace{0.07cm}
       \STATE Compute $G^\pi_{s_{i}} = \displaystyle\sum^{N_{e}}_{j=1} R_{t_{j}}(\pi_{s_{i}})$
    \ENDFOR
    \STATE $G^\pi_{\text{max}} = \text{max}(G^\pi_{s_{1}},G^\pi_{s_{2}},...,G^\pi_{s_{m}})$
    \vspace{0.05cm}
    \STATE $G^\pi_{\text{min}} = \text{min}(G^\pi_{s_{1}},G^\pi_{s_{2}},...,G^\pi_{s_{m}})$
    \vspace{0.05cm}
    \STATE $G^\pi_{\text{scaled}, s_{i}} = \displaystyle\frac{G^\pi_{s_{i}} - G^\pi_{\text{min}}}{G^\pi_{\text{max}} - G^\pi_{\text{min}}}~\forall i \in [1,m]$
    \vspace{0.05cm}
    \STATE $\mathscr{f}_{\text{ranking}}:  G^\pi_{\text{scaled}, s_{i}}\rightarrow T_{tr,s_{i}}~\forall i \in [1,m] $ %
    \STATE Select $k$ out of $m$ sources with $T_{tr,s_{i}} \geq T_{th}$, where $T_{tr,s_{i}}, T_{th} \in [0,1]$ %
\end{algorithmic}
\end{algorithm}%

In Algorithm~\ref{source_ranking}, cumulative return is computed using (\ref{eq:semantic_similarity}) for each source RSU $s_{i}$ with a policy $\pi_{s_{i}}$ and Min-Max normalization is applied to scale cumulative returns $\{G^\pi_{s_{i}}\}^{m}_{i=1}$ into the range of [0,1]. Min-Max normalization based on cumulative return values is defined as 
\setlength{\abovedisplayskip}{7pt}
\setlength{\belowdisplayskip}{7pt}
\begin{align}
    G^\pi_{\text{scaled}, s_{i}} = \displaystyle\frac{G^\pi_{s_{i}} - G^\pi_{\text{min}}}{G^\pi_{\text{max}} - G^\pi_{\text{min}}}, 
    \label{eq:max_min_norm}
\end{align}

\noindent where $G^\pi_{\text{max}}$ and $G^\pi_{\text{min}}$ denote $\text{max}\{G^\pi_{s_{i}}\}^{m}_{i=1}$ and $\text{min}\{G^\pi_{s_{i}}\}^{m}_{i=1}$, respectively.
As such, the scaled cumulative return $G^\pi_{\text{scaled}, s_{i}}$ of a source RSU $s_{i}$ can be associated with its trust value $T_{tr,s_{i}}$, which is used by the target RSU to decide whether the source %
is trustworthy to collaborate with.
We introduce a source ranking strategy $\mathscr{f}_{\text{ranking}}$ in Algorithm~\ref{source_ranking} (line $7$), 
used by the target RSU to sort source RSUs based on their trust values %
and select a subset lying above a specified threshold value $T_{th} \in [0,1]$.
The selection of a threshold to ascertain the trustworthiness of source RSUs can be either %
tolerant with a lower $T_{th}$ value or more stringent with a higher $T_{th}$ value. 
Reputation or trust-based methods in related literature typically use a $[0,1]$ scale to represent trust/reputation values and consider $0.5$ as a neutral value or a threshold~\cite{9003228,9428520}.

\subsubsection{Selective Knowledge Transfer} 
Upon the completion of
source RSUs' selection, the target RSU applies instance TL by collecting samples from $k$ trustworthy source RSUs following their policies $\{\pi_{s_{i}}\}^k_{i=1}$ (i.e., the output of Algorithm~\ref{source_ranking}). In target RSU training, we employ a selective knowledge transfer scheme, called \textit{experience selection}, that selects source samples with high semantic relatedness. Motivated by%
~\cite{tao2021repaint,oh2018self}, 
experience selection %
is defined as
\begin{align}
     Q^*(\REVISED{\mathbf{s}},a) \geq y > Q_{\REVISED{\mathbf{\theta}}}(\REVISED{\mathbf{s}},a),
     \label{eq:related_samples}
\end{align}

\noindent where $Q^*(.)$ represents an optimal $Q$-function and $y = R_t(\REVISED{\mathbf{s}},a) + \gamma\text{max}_{a_{t+1}}Q_{\REVISED{\mathbf{\theta}}}(\REVISED{\mathbf{s_{t+1}}},a_{t+1})$.
The aim of the target RSU is to learn the optimal misbehavior detection policy $\pi^*_{t}$ by obtaining an optimal $Q$-function $Q^*(\REVISED{\mathbf{s}},a)$ for a given state-action pair. 
The expected return achieved by following an optimal policy is always greater or equal to that of any arbitrary behavior policy $\mu$; hence, the expected return $Q_{\mu}(\REVISED{\mathbf{s}},a)$ of any arbitrary behavior policy $\mu$ can serve as a lower bound of the optimal $Q$-value $Q^*(\REVISED{\mathbf{s}},a)$. Lower bound $Q$-learning~\cite{oh2018self} can be expressed as 
\setlength{\abovedisplayskip}{1pt}
\setlength{\belowdisplayskip}{3pt}
\begin{align}
     Q^*(\REVISED{\mathbf{s}},a)  \geq Q_{\mu}(\REVISED{\mathbf{s}},a) = \E_{\mu}[r_{t} + \sum_{k=t+1}^{T} \gamma^{k-t}r_{k}],\label{eq:soft_q_learning}
\end{align}
\noindent 
with convergence guarantees of $Q$-values~\cite{he2016learning}.
The lower bound $Q$-value in (\ref{eq:soft_q_learning}) implies that the estimated $Q_{\REVISED{\mathbf{\theta}}}(\REVISED{\mathbf{s}},a)$ value in (\ref{eq:related_samples}) is lower than the optimal $Q^*(\REVISED{\mathbf{s}},a)$ value. Following lower bound Q-learning, experience selection aims to update $Q_{\REVISED{\mathbf{\theta}}}(\REVISED{\mathbf{s}},a)$ towards the lower bound of $Q^*(\REVISED{\mathbf{s}},a)$ with source samples of high semantic relatedness, i.e., $Q_{\REVISED{\mathbf{\theta}}}(\REVISED{\mathbf{s}},a) \geq y$.

\begin{algorithm}[!ht]
\footnotesize
\caption{Target RSU training with transferred knowledge} %
\label{selective_transfer_DRL}
\begin{algorithmic}[1]
\REQUIRE Load $\pi_{s_{1}}, \pi_{s_{2}},...,\pi_{s_{k}}$ (output of Algorithm~\ref{source_ranking});\\ Initialize experience samples buffer $\tilde{\mathcal{S}}$, replay memory buffer $\mathcal{D}$, action-value function $Q$ with random weights $\theta$ and discount factor $\gamma$ 

\FOR{episode = 1, $M$}
    \STATE Initialize state sequence $\REVISED{\mathbf{s_{1}}}$ %
        \FOR{$t$ = 1, $T$}
            \STATE Pick a random value $rnd \in (0,1)$
            \IF{$\epsilon > rnd$}
                \STATE Select a random action $a_{t} \in \mathcal{A}$ with probability $\epsilon$
            \ELSE
                \STATE Select $a_{t} \in \mathcal{A}$ following policy $\pi_{t}$ %
            \ENDIF
            \STATE Execute $a_{t}$, and observe reward $r_{t}$ and next state $\REVISED{\mathbf{s_{t+1}}}$
            \STATE Collect samples $\mathcal{D} = \{(\REVISED{\mathbf{s_{t}}},a_{t},\REVISED{\mathbf{s_{t+1}}},r_{t})\}$ using $\pi_{t}$
            \STATE Collect samples $\tilde{\mathcal{S}} = \{(\REVISED{\mathbf{\tilde{s}_{t}}},\tilde{a}_{t},\REVISED{\mathbf{\tilde{s}_{t+1}}},\tilde{r}_{t})\}$ using $\pi_{s_i} \forall i \in [1,k]$
            \STATE $\tilde{\mathcal{S}} \leftarrow \tilde{\mathcal{S}} \cup \mathcal{D}$ %
            \STATE Sample random minibatch of transitions $(\REVISED{\mathbf{s_{t}}},a_{t},\REVISED{\mathbf{s_{t+1}}},r_{t})$ from $\tilde{\mathcal{S}}$ %
            \STATE Set $y_{t} = \begin{cases}
                            r_{t}, \qquad \qquad \qquad \qquad \qquad \, \qquad \qquad \text{if}~\REVISED{\mathbf{s_{t+1}}}~\text{terminal}\\    
                            r(\REVISED{\mathbf{s_{t}}},a_{t}) + \gamma \text{max}_{a_{t+1}} Q_{\REVISED{\mathbf{\theta}}} (\REVISED{\mathbf{s_{t+1}}},a_{t+1}), \quad \, \, \, \text{otherwise}%
                    \end{cases}$
            \IF{$Q_{\REVISED{\mathbf{\theta}}}(\REVISED{\mathbf{s_{t}}},a_{t}) \geq y_{t}$} 
                \STATE Select corresponding samples for target model update %
            \ELSE   
                \STATE Remove corresponding transitions $(\REVISED{\mathbf{\tilde{s}_{t}}},\tilde{a}_{t},\REVISED{\mathbf{\tilde{s}_{t+1}}},\tilde{r}_{t})$ from $\tilde{\mathcal{S}}$
            \ENDIF
            \STATE Gradient descent on %
            $(Q_{\REVISED{\mathbf{\theta}}}(\REVISED{\mathbf{s_{t}}},a_{t}) - y_{t} )^2 $ according to (\ref{eq:minimize_loss})%
        \ENDFOR
\ENDFOR
\end{algorithmic}
\end{algorithm} 

The target RSU training with selective knowledge transfer is summarized in Algorithm~\ref{selective_transfer_DRL}. The algorithm proceeds as follows to improve the learning performance of the target RSU with the aid of experience selection. %
The training of the target RSU is divided into episodes. In each time step, the target RSU selects an action $a_{t} \in \mathcal{A}$ either randomly with probability $\epsilon$ or according to the best $Q$-value given by (\ref{eq:eps_greedy}) (lines $4$-$9$). Subsequently, the chosen action $a_{t}$ is executed in the environment, and
the reward $r_{t}$ and the next state $\REVISED{\mathbf{s_{t+1}}}$ are observed (line $10$).  
Assume $k$ trustworthy source RSUs with policies $\{\pi_{s_{i}}\}^{k}_{i=1}$. %
The target RSU can form an experience samples buffer $\tilde{\mathcal{S}}$ by collecting samples following $k$ source policies (line $12$) and computes rewards using its current reward function.
The amount of samples collected into $\tilde{\mathcal{S}}$ from each
trustworthy source RSU $s_{i}$ is equivalent to $\eta_{s_{i}}\cdot\lvert \tilde{\mathcal{S}} \rvert$, with
\setlength{\abovedisplayskip}{5pt}
\setlength{\belowdisplayskip}{1pt}
\begin{align}
    \eta_{s_{i}} =  \frac{T_{tr,s_{i}}}{\sum\limits^{k}_{i=1} T_{tr,s_{i}}}, \quad \forall i \in [1,k],
    \label{eq:sample_proportion}
\end{align}
\noindent 
where $T_{tr,s_{i}} = G^\pi_{\text{scaled}, s_{i}}$ in (\ref{eq:max_min_norm}).
Consequently, source RSUs with higher trust values will transfer more samples, which results in learning more from a set of highly trusted RSUs. 
To trade off exploitation with exploration, target samples are also 
added to $\tilde{\mathcal{S}}$
following the online $Q$-network with $\epsilon$-greedy policy (line $13$).
In the target's model update, parameterized by $\theta$,
training samples are drawn from $\tilde{\mathcal{S}}$ (line $14$), 
and experience selection is applied (lines $16$-$17$) to filter out relevant samples.
Conversely, samples that are not semantically important will be removed from $\tilde{\mathcal{S}}$ (lines $18$-$19$). As such, $\tilde{\mathcal{S}}$ is gradually updated while equally prioritizing the remaining training samples, which results in improved sample efficiency. The target RSU can thus leverage selective knowledge transfer and train its misbehavior detection model by minimizing the loss function,%
\setlength{\abovedisplayskip}{5pt}
\setlength{\belowdisplayskip}{1pt}
\begin{align}
     L(\REVISED{\mathbf{\theta}}) = \displaystyle\frac{1}{2}\sum_t\norm{Q_{\REVISED{\mathbf{\theta}}}(\REVISED{\mathbf{s_{t}}},a_{t}) - y_{t}}^2,
     \label{eq:minimize_loss}
\end{align}

\noindent where target $Q$-values are given by $y_{t} = R(\REVISED{\mathbf{s_{t}}},a_{t}) + \gamma\text{max}_{a_{t+1}}Q_{\REVISED{\mathbf{\theta}}}(\REVISED{\mathbf{s_{t+1}}},a_{t+1})$. %
Moreover, the selection of samples with $Q_{\REVISED{\mathbf{\theta}}}(\REVISED{\mathbf{s_{t}}},a_{t}) \geq y_{t}$ implies that learning updates in (\ref{eq:minimize_loss}) are encouraged towards
the lower bound of the optimal $Q^*(\REVISED{\mathbf{s}},a)$ value. Consequently, experience selection updates the misbehavior detection policy $\pi_{t}$ towards the optimal policy $\pi^*_{t}$.

Although experience selection improves sample efficiency, transferred knowledge from source instances may introduce bias owing to differences in data distribution or variability in misbehavior attack patterns.
Therefore, to circumvent such bias, experience selection is utilized only in early training stages and, subsequently, target RSU training switches to traditional DQN learning.
Algorithm~\ref{selective_transfer_DRL} for traditional DQN operates similarly but without sampling from $\tilde{\mathcal{S}}$ (lines $13$-$14$) and with no experience selection (lines $16$-$20$). 
In this case, the target RSU samples a minibatch of transitions from replay memory buffer $\mathcal{D}$ with training samples from the target environment, and those samples are directly used to train the misbehavior detection model by minimizing  (\ref{eq:minimize_loss}).

\section{Experiments}
\label{sec:experiments}

This section describes the experimental scenarios and the setup used to verify the validity and effectiveness of the proposed knowledge transfer approach for collaborative misbehavior detection. We leverage the open-source VeReMi dataset for experimentation~\cite{kamel-veremi-20}.

\subsection{Dataset Description and Pre-processing}

The VeReMi dataset has been generated considering
$19$ different misbehavior attack types
with
each attack type 
having a separate dataset. An attack dataset is made of a collection of log files, each one 
generated containing BSM data 
exchanged with
neighboring vehicles over 
their
entire trajectory. Each attack dataset contains a ground truth file that has recorded the observed behavior of all participating vehicles. BSMs constitute a three-dimensional vector for the position, speed, acceleration, and heading angle parameters. The ratio between misbehaving and genuine vehicles 
in VeReMi
is 30\%:70\% for all the attack simulations. %
In Appendix~\ref{sec:apx_attacks}, we briefly describe selected misbehavior attack types relevant to our study.

Based on feature analysis and data pre-processing, we selected six fields, i.e., timestamp, pseudo-identity, position, speed, acceleration, and heading angle, from a BSM payload as the relevant features for misbehavior detection. To reduce dimensionality, a scalar representation of the position, speed, acceleration, and heading parameters is computed using the Euclidean norm. In VeReMi, the ground truth data do not contain labels; hence, a label for each data point was generated by comparing the ground truth against the actual transmitted value recorded in the log file of each vehicle. An attack label $1$ was added if the transmitted data diverged from the ground truth; otherwise, label $0$ was added for the genuine data.

\subsection{Examined Scenarios} 
\label{ssec:examined_scenarios}
We conduct experiments on a set of simulation
scenarios, listed in Table~\ref{tab:attack_scenarios}, to assess the proposed knowledge transfer approach. \REVISED{We provide empirical evidence of the benefits of TL through simulated scenarios for collaborative misbehavior detection in unpredictable and untrusted vehicular environments.}  
The considered scenarios aim to cover diversified aspects of collaborative misbehavior detection by means of knowledge transfer between RSUs. \REVISED{By leveraging knowledge from source RSUs with high semantic relatedness, the target RSU can rapidly adapt to new misbehavior patterns, even in dynamic and volatile conditions.}
Vehicular mobility
renders indispensable the study of 
such scenarios that may arise in a collaborative misbehavior detection setup.
With that aim, we exploit the attack variants present in VeReMi. For each specific scenario, a different set of misbehavior attacks is selected, aiming to provide sufficient coverage of the available attacks in VeReMi. 
Next, we elaborate on the considered scenarios.
\begin{table}[!t]
\caption{Examined Scenarios\label{tab:attack_scenarios}}
\centering
\resizebox{0.49\textwidth}{!}{
\begin{tabular}{|c|l|}
\hline
 \textbf{Scenario} & \multicolumn{1}{c|}{\textbf{Objective}}\\
\hline
SC1 & Detect future misbehaviors of the same type\\
\hline
 SC2  & Detect unseen/unknown misbehaviors of similar type\\
\hline
SC3 &  Detect partially observable misbehaviors\\
\hline
\end{tabular}}
\end{table}
\subsubsection{SC1} The first
scenario captures
the requirement of detecting future misbehaviors of the same type.
In SC1, a misbehaving vehicle can permeate an attack across multiple geographic locations 
across its trajectory, 
resulting in some RSUs experiencing the attack earlier
in time (sources) compared to others (targets). SC1 arises in situations where RSU deployments are sparsely distributed in vehicular setups.
In this case, the target RSU 
relies on collaborative misbehavior detection to proactively detect future attacks of the same type.

By exploiting the evolving attack patterns%
, we split an attack dataset based on the timestamp field (ascending order) and assign a similar amount of training samples to each source RSU. To motivate knowledge transfer, the number of training samples allocated to a source RSU is proportionally %
higher than that of the target RSU. %
It is noted that the allocated samples 
per RSU
preserve the ratio between misbehaving and genuine classes as set in the entire dataset.   %

\subsubsection{SC2} In this scenario, the target RSU aims to acquire knowledge from source RSUs to detect an unseen/unknown attack. Such situations may encounter in practice due to blind spots and occlusions caused by vehicular infrastructure or moving vehicles. This inevitably results in limited situational awareness in some RSUs (targets) which may not experience certain attack types. In this case, the target RSU seeks relevant knowledge from source RSUs with expertise in detecting attacks of similar type.
We assume that source RSUs are capable of detecting %
a wider range of misbehavior attack types. %
On the contrary, the target RSU is trained to identify a narrower set 
of attacks %
compared to
source RSUs. %
In such cases, the target RSU aims to leverage knowledge transfer from source RSUs to identify similar new misbehavior attacks.

The VeReMi dataset includes a set of misbehavior attack types that have been generated by executing a combination of multiple attacks at once (as detailed in Appendix~\ref{sec:apx_attacks}). We hereby evaluate two cases within this scenario, each involving different denial-of-service (DoS) variants:
\begin{enumerate}[label=\roman*)]
    \item Each source RSU is trained on a mix of DoS, DoS Random and DoS Random Sybil attacks. The target RSU is trained on a mix of DoS and DoS Random attacks, and leverages source knowledge to detect the unseen DoS Random Sybil attack.
    \item Each source RSU is trained on a mix of DoS, DoS Disruptive and DoS Disruptive Sybil attacks. The target RSU is trained on a mix of DoS and DoS Disruptive attacks, and leverages source knowledge to detect the unseen DoS Disruptive Sybil attack.
\end{enumerate}%

\subsubsection{SC3} In this scenario, 
source RSUs are trained with different feature vectors resulting in partial observability of the attack space. Such cases arise in practice when
RSUs have diverse computational capabilities, including different processing and buffer sizes, 
limiting their training only on specific attack types.
This heterogeneity may hinder an RSU’s ability to process multiple/high-dimensional feature vectors effectively.
Thus, the target RSU needs to select source RSUs based on relevant feature vectors to detect an array of misbehaviors and enhance the effectiveness of collaborative detection.
We assume that each source RSU $s_i$ is trained with feature vector $\REVISED{\mathbf{\phi_i}}$ on a targeted misbehavior attack type. 
Similarly, the target RSU is trained on a specific misbehavior attack type with a feature vector $\REVISED{\mathbf{\phi_t}}$ with dimensionality $\rho_{\REVISED{\mathbf{\phi_t}}}$$<$$\rho_{\REVISED{\mathbf{\phi_i}}}$, $\forall i$. Then, the target RSU can leverage source knowledge by combining feature vectors, $\bigcup \REVISED{\mathbf{\phi_i}}$,
to detect %
a wider range of misbehaviors.
We consider position- and speed-related attack variants in VeReMi, distinguishable by various feature values embedded in BSMs. Similarly to SC2, we evaluate two cases:
\begin{enumerate}[label=\roman*)]
    \item Each source RSU is individually trained on
    Constant Position Offset, Random Position, and Random Position Offset attacks. The target RSU is trained on Constant Position attack, and leverages source knowledge to detect all position-related attack variants.
    \item Each source RSU is individually trained on
    Constant Speed Offset, Random Speed, and Random Speed Offset attacks. The target RSU is trained on Constant Speed attack, and leverages source knowledge to detect all speed-related attack variants.
\end{enumerate}%

\subsection{Experimental Setup} 
The experiments were performed on a %
server equipped with Intel(R) Xeon(R) Gold 6230 CPUs~@~2.10GHz with 192 GB RAM and four NVIDIA GeForce RTX 2080 Ti GPUs. The components of the DRL-MDS (Fig.~\ref{fig:drl_mds_rsu}) are implemented in Python using Tensorflow-GPU 
as the backend. %
Considering the amount of data samples available per misbehavior type in VeReMi, we resort to a number of four RSUs in all scenarios, with %
three source RSUs and a single target RSU.
Source RSUs comprise two genuine RSUs and an adversary-influenced (i.e., malicious) RSU. The malicious RSU is trained using a single type of adversarial 
attack at a time, either label flipping or policy induction attack.
In this way, the level of impact on knowledge transfer from each adversarial attack type can be assessed separately. The learning rate $\alpha$ and discount factor $\gamma$ in DRL-MDS are set to $0.001$ and $0.995$, respectively. 
Both $\tilde{\mathcal{S}}$ and $\mathcal{D}$ sizes at the target RSU are set to $50000$.

In the label-flipping attack, the parameter $\zeta$ is set to $0.05$, which is equivalent to 
randomly selecting $5\%$ of misbehaving vehicles from a source RSU and flip the labels of the corresponding BSMs to create a malicious RSU. In the policy induction attack, the adversarial policy $\pi_{adv}$ is obtained based on an inverted reward function $r'$ compared to (\ref{eq:rew_func}).
The $r'$ generates exact opposite reward values for the malicious RSU to maximize FNs and FPs. To obtain $\pi_{adv}$, an adversarial DQN model is trained with $r'$ on a small portion of training samples extracted from a victim source RSU's environment.
With $\pi_{adv}$, the policy induction attack (as in Definition~\ref{def:adv_policy}) is launched by generating perturbed samples during the training of the victim source RSU to create a malicious RSU. 

\section{Performance Evaluation}
\label{sec:perf_evaluation}

This section presents the learning performance of DRL-MDS agents
for the scenarios described above using VeReMi. Next, the resulting detection performance achieved via knowledge transfer is assessed for different misbehavior attack types.

\subsection{Learning Performance}
To provide a comprehensive assessment, we evaluate the learning performance of misbehavior detection agents in \textit{i}) the source RSUs with trained policies and \textit{ii}) the target RSU with and without knowledge transfer. 

\subsubsection{Source RSUs} 
To assess the contribution of source RSUs towards knowledge transfer, we measure the average reward gains per RSU during the interactions with their own environments. 
In each scenario, a source RSU is set to run for 500 training episodes, 
and the learning performance is measured in terms of the cumulative reward values obtained following the reward function in (\ref{eq:rew_func}). %

\begin{figure}[!t]
\captionsetup[subfigure]{labelformat=empty,skip=0pt}
\centering
\subfloat[]{\includegraphics[width=.25\textwidth]{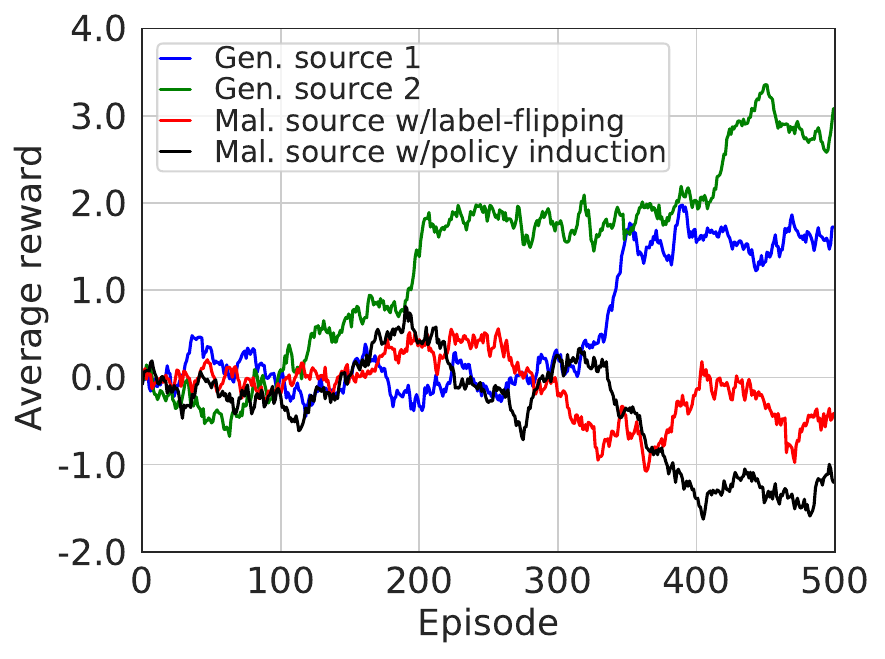}%
\label{fig:randposoffset_src_training}}
\subfloat[]{\includegraphics[width=.25\textwidth]{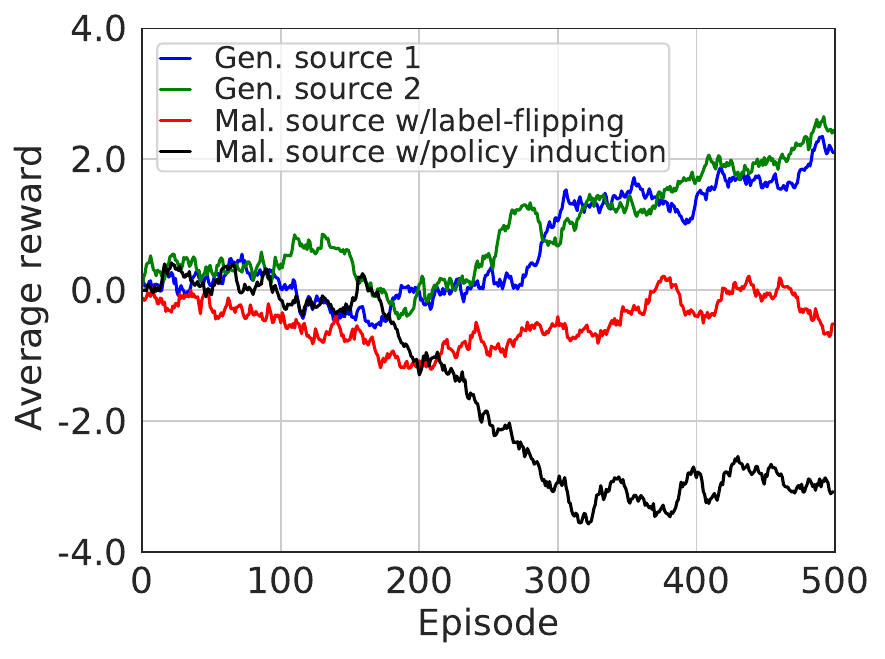}%
\label{fig:dos_rnd_sybil_src_training}}
\caption{Learning performance of source RSUs in SC1 for (Left) Random Position and (Right) Random Position Offset misbehavior attack types.}
\label{fig:sc1_source_policies_training}
\end{figure}

\begin{figure}[!t]
\captionsetup[subfigure]{labelformat=empty,skip=0pt}
\centering
\subfloat[]{\includegraphics[width=.25\textwidth]{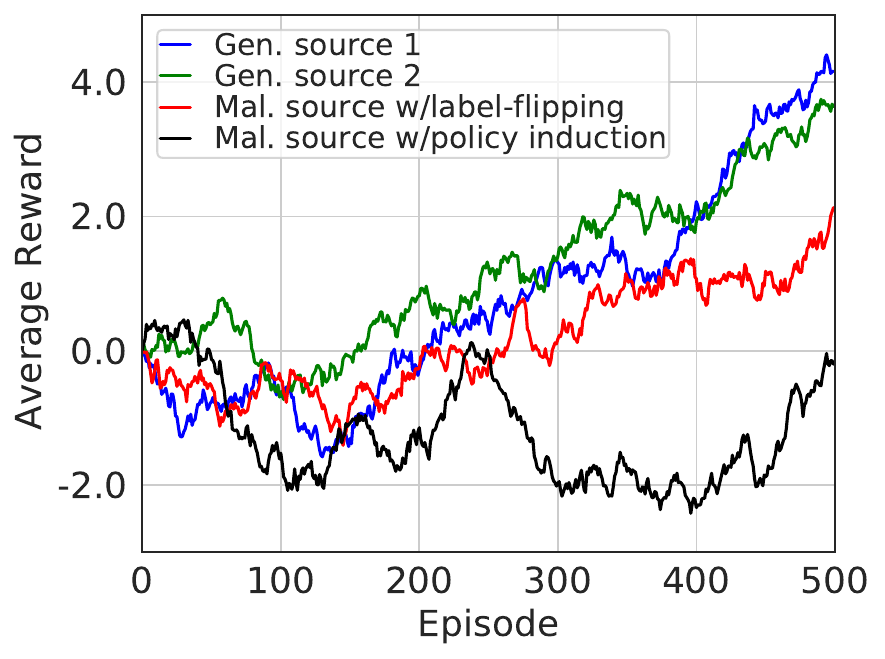}%
\label{fig:dos_rnd_sybil_src_training}}
\subfloat[]{\includegraphics[width=.25\textwidth]{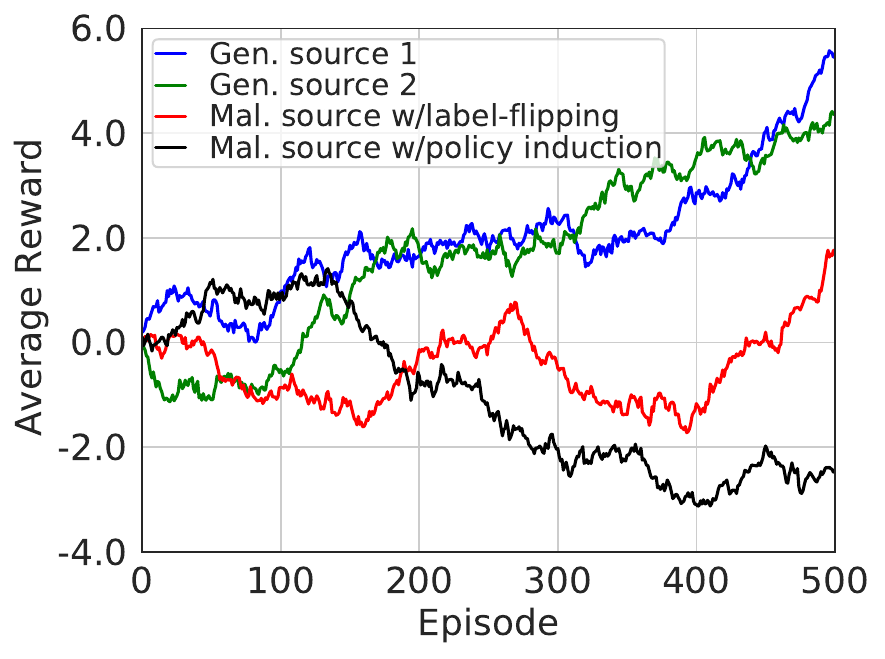}%
\label{fig:dos_disrupt_sybil_src_training}}
\caption{Learning performance of source RSUs in SC2 for (Left) a combination of DoS, DoS Random and DoS Random Sybil; (Right) a combination of DoS, DoS Disruptive and DoS Disruptive Sybil misbehavior attack types.}
\label{fig:sc2_source_policies_training}
\end{figure}

\begin{figure}[!t]
\captionsetup[subfigure]{labelformat=empty,skip=0pt}
\centering
\subfloat[]{\includegraphics[width=.25\textwidth]{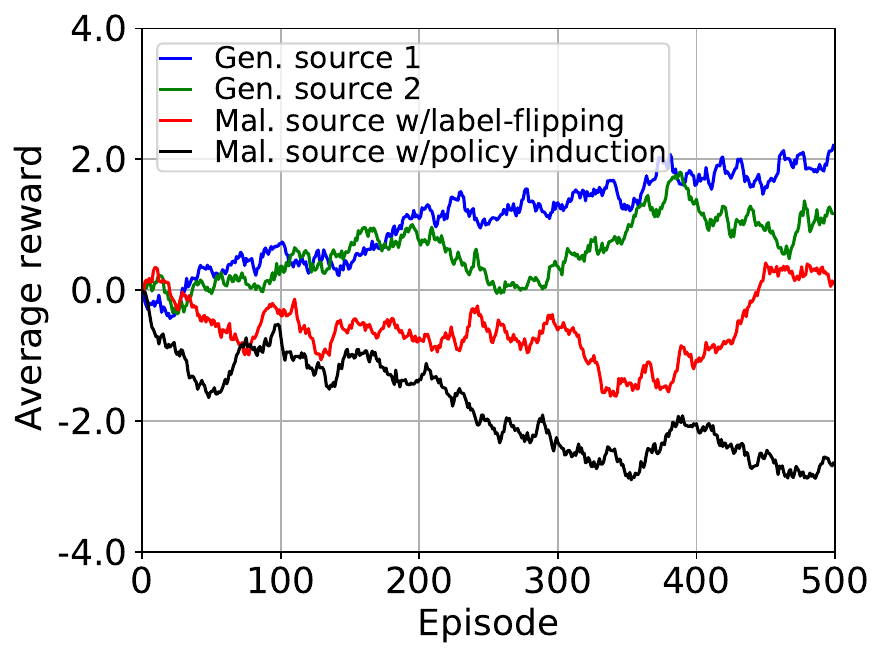}%
\label{fig:sc3_position_related_src_training}}
\subfloat[]{\includegraphics[width=.25\textwidth]{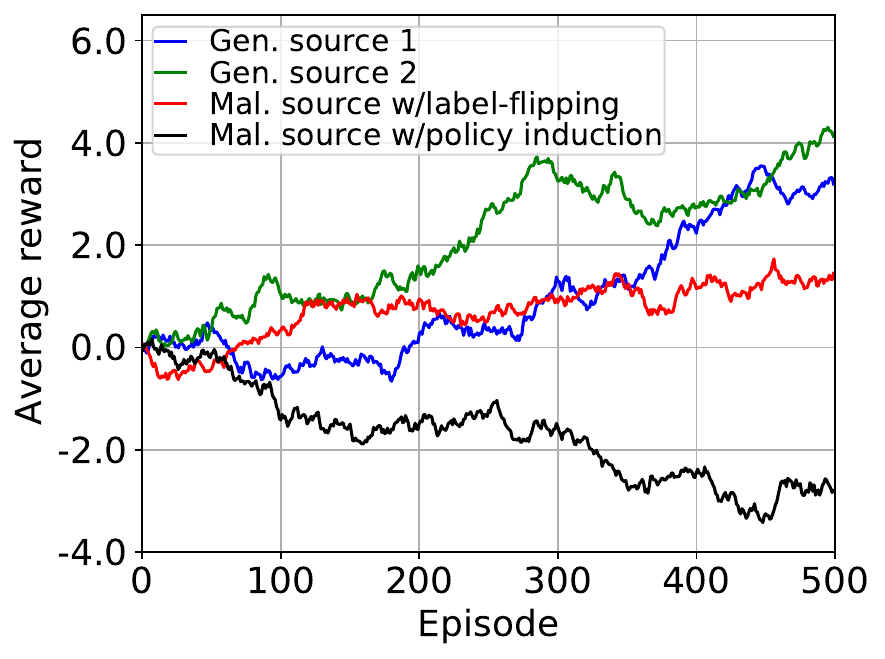}%
\label{fig:sc3_spd_related_src_training}}
\caption{Learning performance of source RSUs in SC3 for (Left) position variants with Constant Position Offset (Gen. source 1), Random Position (Gen. source 2), and Random Position Offset (Mal. sources) misbehaviors; (Right) speed variants with Constant Speed Offset (Gen. source 1), Random Speed (Gen. source 2), and Random Speed Offset (Mal. sources) misbehaviors.}
\label{fig:sc3_source_policies_training}
\end{figure}

Fig.~\ref{fig:sc1_source_policies_training} illustrates the learning performance of source RSUs for two position-related misbehavior types considered for SC1. As can be visually comprehended, genuine source RSUs accumulate significantly higher average rewards compared to malicious source RSUs, which are detrimentally influenced by label-flipping and policy induction attacks. %
Furthermore, the negative impact on the victim source RSU from the policy induction attack 
is greater than that of the label-flipping attack. 
This can be attributed to
the effectiveness of policy induction attack which injects crafted inputs at each training step 
towards the adversary's goal. This, in turn,
results in misclassifying malicious samples as genuine ones with increased false alarms.  
Similar behavior can be observed in Fig.~\ref{fig:sc2_source_policies_training} for DoS-related misbehavior variants considered in SC2. In this case, each source RSU is trained on a combination of misbehavior attack types, providing the capability to detect additional misbehavior attack types compared to SC1. In Fig.~\ref{fig:sc3_source_policies_training}, the learning performance of source RSUs in SC3 for both position- and speed-related misbehavior variants is shown. Similarly to SC1 and SC2, genuine sources perform significantly better than malicious ones, while
the pernicious impact of policy induction attack compared to label-flipping becomes apparent.

\subsubsection{Target RSU} 

As \UPDATED{shown} in Algorithm~\ref{source_ranking}, 
based on the available source policies, the target RSU selects a subset of source RSUs using $T_{th}$ to realize knowledge transfer. As opposed to conventional predetermined thresholds (Sec.~\ref{ssec:trust}), the proposed trust evaluation supports dynamic thresholding to handle 
unpredictable and untrusted vehicular environments,
such as the erratic behavior of RSUs due to adversarial attacks. %
For instance, several previous works use $0.5$ as the neutral threshold value. 
In our case, we select $0.5$ and $0.8$ for $T_{th}$ to assess the impact of different threshold values on knowledge transfer without loss of generality.  %

\begin{figure}[!t] %
\captionsetup[subfigure]{labelformat=empty,skip=0pt}
\centering
\subfloat[]{\includegraphics[width=0.24\textwidth,keepaspectratio]{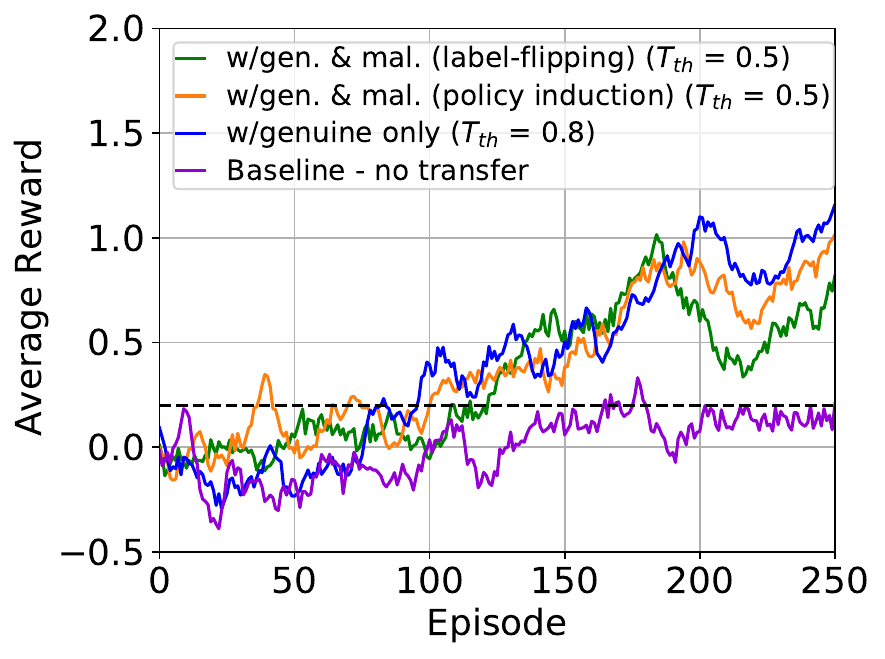}}%
\label{fig:randpos_tgt_training}
\subfloat[]{\includegraphics[width=0.24\textwidth,keepaspectratio]{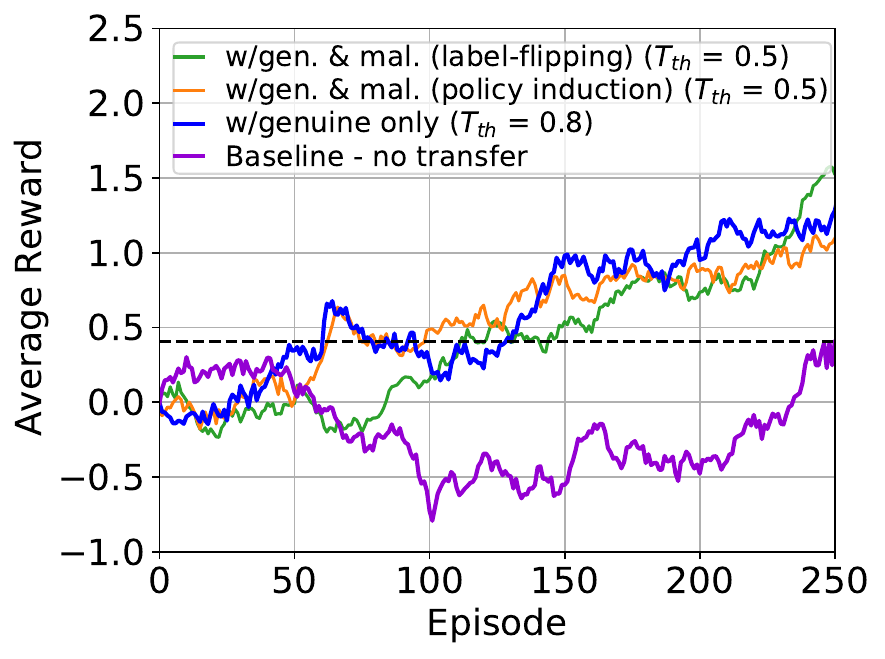}}%
\label{fig:rndpos_offset_tgt_training}
\caption{Learning performance of the target RSU, averaged across three runs, in SC1 for (Left) Random Position and (Right) Random Position Offset misbehavior attack types.}
\label{fig:rndpos_rndposoffset_target_training}
\end{figure}

\begin{figure}[!t] %
\captionsetup[subfigure]{labelformat=empty,skip=0pt}
\centering
\subfloat[]{\includegraphics[width=0.24\textwidth,keepaspectratio]{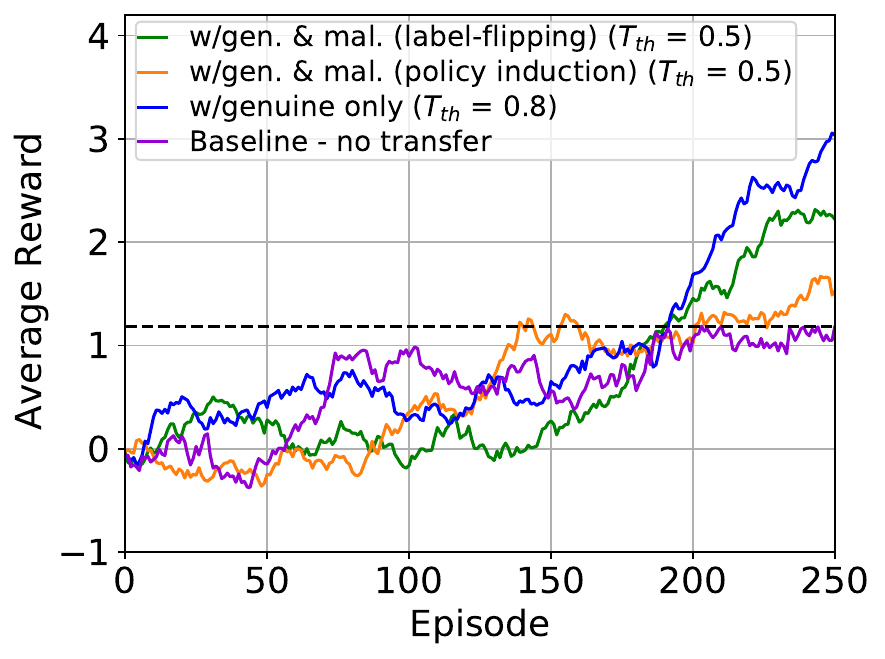}}%
\label{fig:dos_dosrnd_dosrndsybil_target_training}
\subfloat[]{\includegraphics[width=0.24\textwidth,keepaspectratio]{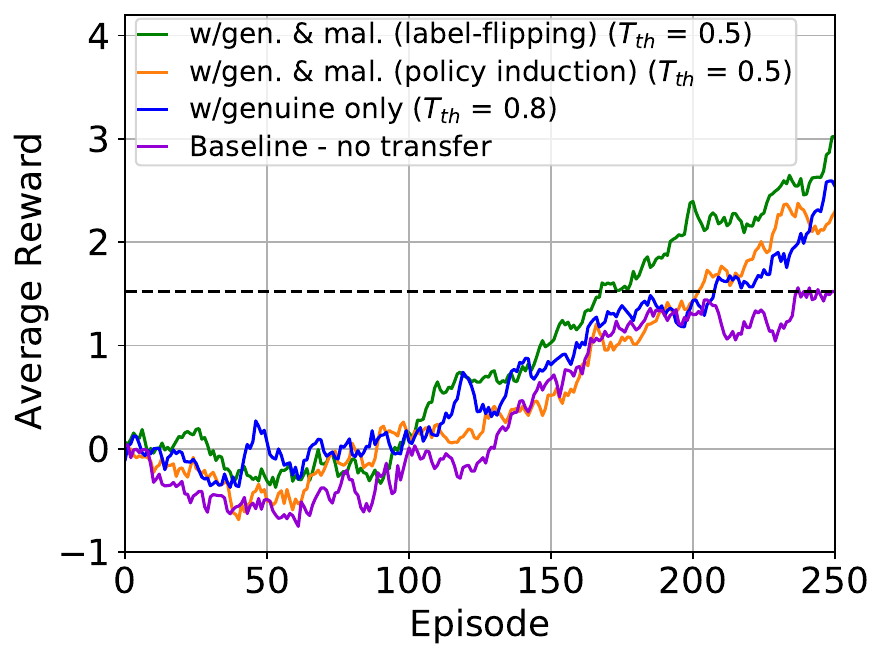}}%
\label{fig:dos_dosdist_dosdistsybil_target_training}
\caption{Learning performance of the target RSU, averaged across three runs, in SC2 for (Left) a combination of DoS and DoS Random; (Right) a combination of DoS and DoS Disruptive misbehavior attack types. %
}
\label{fig:dosrndsybil_dosdistsybil_target_training}
\end{figure}

\begin{figure}[!t] %
\captionsetup[subfigure]{labelformat=empty,skip=0pt}
\centering
\subfloat[]{\includegraphics[width=0.24\textwidth,keepaspectratio]{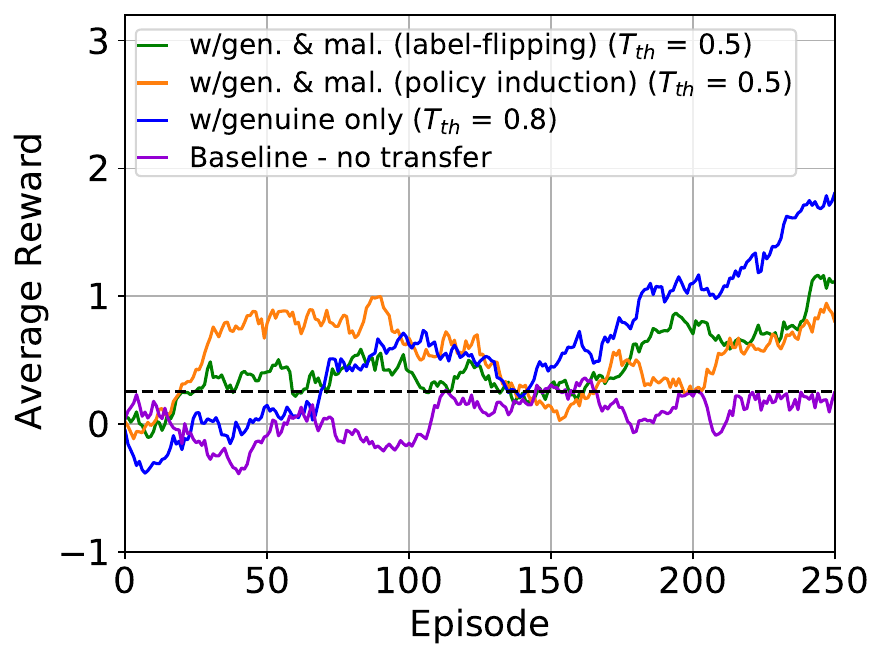}}%
\label{fig:sc3_position_related_target_training}
\subfloat[]{\includegraphics[width=0.24\textwidth,keepaspectratio]{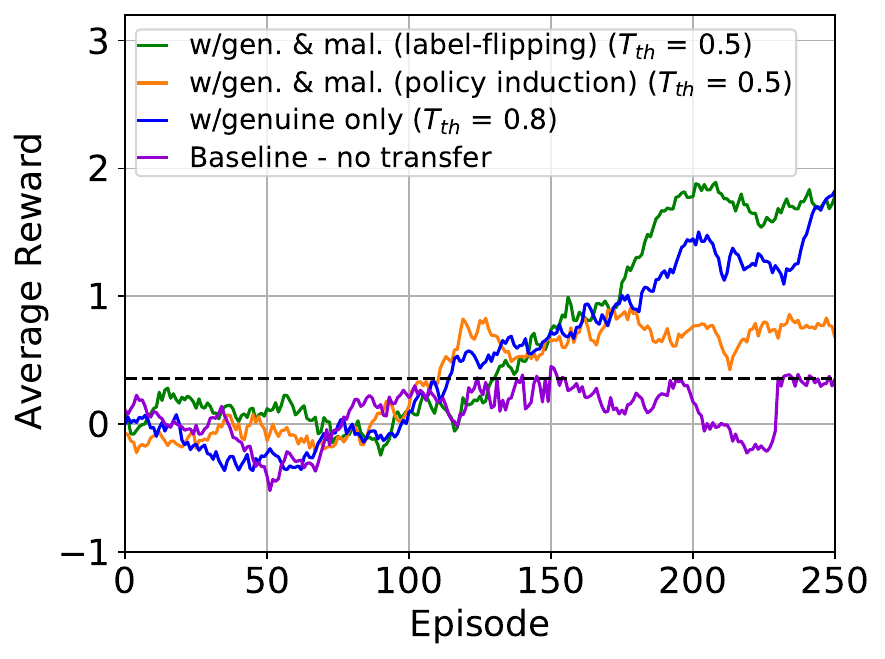}}%
\label{fig:sc3_speed_related_target_training}
\caption{Learning performance of the target RSU, averaged across three runs, in SC3 for (Left) position-related and (Right) speed-related misbehavior variants. %
}
\label{fig:sc3_position_speed_target_training}
\end{figure}

Figs.~\ref{fig:rndpos_rndposoffset_target_training}, \ref{fig:dosrndsybil_dosdistsybil_target_training}, and \ref{fig:sc3_position_speed_target_training} depict the learning performance of the target RSU in SC1, SC2, and SC3, respectively. 
We report average rewards for knowledge transfers involving different sources by calculating the average across three runs. The case of no transfer between source and target RSUs (i.e., tabula rasa learning) is also evaluated as a baseline. \UPDATED{In addition, the black dashed lines in Figs.~\ref{fig:rndpos_rndposoffset_target_training}--\ref{fig:sc3_position_speed_target_training} represent the best return reward of the baseline scheme with tabula rasa learning. These lines demonstrate the training time reduction achieved under each transfer at the target RSU to reach a certain performance level. Moreover, Table~\ref{tab:training_time_reduction} presents specific information regarding the reduction in training time achieved through knowledge transfer, detailing the required number of training episodes and the corresponding wall-clock time.}
Fig.~\ref{fig:rndpos_rndposoffset_target_training} illustrates average rewards accumulated by the target RSU for a specific misbehavior with knowledge transferred from the source RSUs (shown in Fig.~\ref{fig:sc1_source_policies_training}). It can be observed that our selective knowledge transfer approach significantly enhances the learning performance compared to tabula rasa.
Specifically, 
the target RSU achieves a higher cumulative return and reaches asymptotic performance earlier than in the baseline. 

In addition, Fig.~\ref{fig:rndpos_rndposoffset_target_training} 
reveals the impact of different threshold values on learning performance.
When a tolerant $T_{th}$ = $0.5$ value is set, both genuine and malicious (label-flipping and policy induction) source RSUs feature in the transfer. With a stringent $T_{th}$ = $0.8$, collaboration stems only from 
genuine source RSUs. %
Interestingly, even though malicious source RSUs are involved with $T_{th}$ = $0.5$, 
the sample contribution in (\ref{eq:sample_proportion}) and the experience selection in (\ref{eq:related_samples}) ascertain positive transfer, resulting in similar performance as in $T_{th}$ = $0.8$.

\begin{table*}[t!]
\renewcommand{\arraystretch}{1.3}
\caption{Summary of training time and the best return reward at the target RSU under knowledge transfer}
\begin{threeparttable}
\label{tab:training_time_reduction}
\centering
\resizebox{1.0\textwidth}{!}{
\begin{tabular}{|c|l|c|c|c|c|c|c|c|c|c|}
    \hline
    \multirow{2}{*}
   {\parbox{2cm}{\centering \textbf{Examined} \\ \textbf{Scenario}}}
   & \multirow{2}{*}{\textbf{Misbehavior Type(s)}} & \multirow{2}{*}{\textbf{Target Return}} & \multicolumn{4}{|c|}{\textbf{Training time in episodes (reduction \%)}} & \multicolumn{4}{|c|}{\textbf{Training time in wall-clock time (mins) (reduction \%)}} \\ \cline{4-7} \cline{8-11} 
    & & & \textbf{Baseline} & \textbf{TF\tnote{1}} & \textbf{TF\textsuperscript{2}} & \textbf{TF\textsuperscript{3}} & \textbf{Baseline} & \textbf{TF\tnote{1}} & \textbf{TF\textsuperscript{2}} & \textbf{TF\textsuperscript{3}} \\ \hline
    \multirow{2}{*}{SC1} & Random Position & 0.2005 & 250  & \makecell[c]{122 (51\%) \\ (best return: 0.8173)}  & \makecell[c]{100 (60\%)\\(best return: 1.0156)}   & \makecell[c]{96 (62\%)\\ (best return: 1.1575)} & 83.32 & 31.33 (62\%) & 26.68 (68\%)  & 24.35 (71\%)  \\\cline{3-11}
    \cline{2-11}
    & Random Position Offset & 0.4092 & 250  & \makecell[c]{144 (42\%)\\ (best return: 1.5233)}   & \makecell[c]{98 (61\%) \\ (best return: 1.0960)}  & \makecell[c]{128 (49\%)\\ (best return: 1.1574)} & 82.93 & 36.83 (56\%) & 26.14 (68\%) & 32.02 (61\%) \\\cline{3-11}
    
    \hhline{===========}
    \multirow{2}{*}{SC2} & DoS Random Sybil & 1.1878 & 250  & \makecell[c]{190 (24\%) \\ (best return: 2.2230)}  & \makecell[c]{201 (20\%)\\ (best return: 1.5293)} & \makecell[c]{192 (23\%) \\ (best return: 3.0571)} & 84.57 & 50.16 (41\%) & 52.29 (38\%)  & 50.68 (40\%) \\\cline{3-11}
    \cline{2-11}
    & DoS Disruptive Sybil & 1.5235 & 250  & \makecell[c]{168 (33\%) \\(best return: 3.0216)}   & \makecell[c]{202 (19\%)\\ (best return: 2.3775)}  & \makecell[c]{208 (17\%)\\ (best return: 2.5939)}  & 85.89 & 44.23 (48\%)  & 53.12 (38\%) & 55.14 (36\%) \\\cline{3-11}
    
    \hhline{===========}
    \multirow{2}{*}{SC3} & Position variants* &  0.2548 & 250  & \makecell[c]{162 (35\%)\\ (best return: 1.1656)}  & \makecell[c]{166 (34\%)\\(best return: 0.7982)} & \makecell[c]{142 (43\%)\\ (best return: 1.8103)} & 79.18  & 39.37 (50\%)  & 40.35 (49\%)  & 36.20 (54\%) \\\cline{3-11}
    
    \cline{2-11}
    & Speed variants** & 0.3541  & 250  & \makecell[c]{130 (48\%) \\ (best return: 1.7811)}   & \makecell[c]{111 (55\%)\\ (best return: 0.6789)}  & \makecell[c]{114 (54\%) \\ (best return: 1.8242)} & 80.59  & 32.19 (60\%)  & 27.49 (66\%) & 28.65 (64\%) \\\cline{3-11}
    \hline
\end{tabular}}
\begin{tablenotes}\footnotesize
       \item {TF\textsuperscript{1}: w/genuine+malicious (label-flipping) ($T_{th}$ = $0.5$)}
       \item {TF\textsuperscript{2}: w/genuine+malicious (policy induction) ($T_{th}$ = $0.5$)}
       \item {TF\textsuperscript{3}: w/genuine only ($T_{th}$ = $0.8$)}\\
       \item  {\textsuperscript{*} Constant Position Offset, Random Position, and Random Position Offset}
       \item  {\textsuperscript{**} Constant Speed Offset, Random Speed, and Random Speed Offset}
\end{tablenotes}
\end{threeparttable}
\vspace*{-0.5cm}
\end{table*}

Fig.~\ref{fig:dosrndsybil_dosdistsybil_target_training} demonstrates 
the learning performance of the target RSU
with
unseen misbehavior attack variants. 
For the two cases in SC2, the target RSU acquires expert knowledge from sources to detect unseen DoS Random Sybil and DoS Disruptive Sybil misbehaviors.
In contrast, the baseline with tabula rasa 
does not possess prior knowledge of those two unseen variants. %
The learning curves indicate a superior performance of the target RSU compared
to the baseline.
This can be attributed to the fact that collaborative knowledge transfer from source RSUs (Fig.~\ref{fig:sc2_source_policies_training}) imparts the necessary expertise to detect unseen misbehaviors.
Similar to SC1 (Fig.~\ref{fig:rndpos_rndposoffset_target_training}), our approach exhibits consistent learning performance irrespective of the 
$T_{th}$ value. %
Note that DoS misbehaviors in VeReMi are high-volume attacks, resulting in the frequent learning of attack patterns by both source RSUs and the target RSU. Thus, 
the target RSU accumulates a relatively higher total average reward for a similar number of training episodes compared to Fig.~\ref{fig:rndpos_rndposoffset_target_training}.

Fig.~\ref{fig:sc3_position_speed_target_training} displays the learning performance of the target RSU with partial observability of the attack space.
For the two cases in SC3, the target RSU assimilates knowledge from source RSUs (Fig. \ref{fig:sc3_source_policies_training}) to expand the detection horizon to a wider range of misbehaviors. As such, a source RSU learns a specific misbehavior effectively using high-dimensional feature vectors, and then the acquired knowledge is distilled to the target RSU via selective knowledge transfer. %
In particular, each source RSU is trained with a high-dimensional feature vector comprising position, speed, acceleration, and heading angle parameters. The target RSU is trained with the position feature for the Constant Position attack and with the speed feature for the Constant Speed attack.
As shown in Fig.~\ref{fig:sc3_position_speed_target_training}, the target RSU in the baseline scheme achieves notably lower asymptotic performance compared to the target RSU with knowledge transfer. This difference can be attributed to the increased false alarms resulting from the partial observability of the attack space when using low-dimensional feature vectors. To circumvent this, knowledge distillation from sources with multiple/high-dimensional feature vectors can be applied to enhance
the effectiveness of collaborative misbehavior detection.
Note that our approach exhibits 
comparable learning performance for both $T_{th}$ values, similar to SC1 (Fig. \ref{fig:rndpos_rndposoffset_target_training}) and SC2 (Fig. \ref{fig:dosrndsybil_dosdistsybil_target_training}).

\subsection{Detection Performance}
\label{ssec:misdet_performance}
Detection performance is assessed in \CHANGED{all} scenarios in terms of \textit{Accuracy} (A), \textit{Precision} (P), \textit{Recall} (R), and \textit{F-score} (F), 
\setlength{\abovedisplayskip}{7pt}
\setlength{\belowdisplayskip}{7pt}
\begin{align}
    \text{A} & = \frac{TP+TN}{TP+TN+FP+FN},
    \label{eq:1}\\
   \text{P} & = \frac{TP}{TP+FP},
    \label{eq:2}\\
   \text{R} & = \frac{TP}{TP+FN},
   \label{eq:3} \\
   \text{F} & = 2\frac{RP}{R + P},
   \label{eq:4}
\end{align}
\noindent \CHANGED{respectively,} by considering both genuine and misbehavior classes for each misbehavior type in VeReMi.
\noindent The accuracy in (\ref{eq:1}) indicates the ratio of all correct predictions to the total number of considered input samples. The higher precision values in (\ref{eq:2}) indicate low FP rates, whereas higher recall values in (\ref{eq:3}) indicate low FN rates. The F-score in (\ref{eq:4}) provides a harmonic mean between precision and recall, which is used when FPs and FNs are vital. Therefore, a higher F-score implies better performance in our examined scenarios.   

Table~\ref{tab:misdet_perf_results} presents the performance results obtained from our comprehensive analysis of collaborative misbehavior detection with the selective knowledge transfer approach. Results show that misbehavior detection using knowledge transfer yields high effectiveness with a very high F-score of $0.98$ in SC1 under both $T_{th}$ values of $0.5$ and $0.8$, while correctly identifying misbehaviors with low rates of FPs and FNs. It should be highlighted that the baseline scheme with tabula rasa also achieves a significantly high F-score of $0.91$. This is due to the \CHANGED{fact} that tabula rasa learning obtains sufficient knowledge during training to detect future misbehaviors of the same type. Although the results are reported only for two misbehavior types in SC1, similar performance levels were observed for other misbehavior types in VeReMi.

Numerical results 
for SC2 demonstrate
effective identification of unseen attacks with knowledge transfer, achieving an F-score of $0.71$ under both $T_{th}$ values. Additionally, recall values approaching $1.0$ further \CHANGED{elucidate}
that \CHANGED{such non-anticipated}
attacks can be successfully detected with a very low rate of FNs. As shown in Table~\ref{tab:misdet_perf_results}, the baseline scheme is ineffective in detecting unseen attacks and achieves very low F-scores of $0.49$ and $0.48$ in contrast to knowledge transfers when encountering DoS Random Sybil and DoS Disruptive Sybil misbehaviors, respectively. This validates that the target RSU enhances its situational awareness and detects \CHANGED{non-anticipated} misbehaviors %
by acquiring %
knowledge from source RSUs. %

\begin{table*}[t!]
\caption{Performance comparison for collaborative misbehavior detection in each scenario}
\vspace*{-0.2cm}
\label{tab:misdet_perf_results}
\centering
\resizebox{1.0\textwidth}{!}{
\begin{tabular}{|c|l|l|c|c|c|c|}
    \hline
    \textbf{Examined Scenario} & \textbf{Misbehavior Type(s)} & \textbf{Knowledge Transfer} & \textbf{Accuracy} & \textbf{Precision} & \textbf{Recall} & \textbf{F-score} \\ \hline
    \multirow{8}{*}{SC1} & \multirow{4}{*}{Random Position} & w/genuine+malicious (label-flipping) ($T_{th}$ = $0.5$) & 0.9928  & 0.9778  & 0.9983  & 0.9879  \\\cline{3-7}
    & & w/genuine+malicious (policy induction) ($T_{th}$ = $0.5$) & 0.9933 & 0.9789  & 0.9987  & 0.9887  \\\cline{3-7}
    & & w/genuine only ($T_{th}$ = $0.8$) & 0.9920 & 0.9756 & 0.9977 & 0.9865  \\\cline{3-7}
    & & Baseline (no transfer) & 0.9497 & 0.8978 & 0.9362  & 0.9166 \\\cline{3-7}
    \hhline{|~======}
    
    & \multirow{4}{*}{Random Position Offset} & w/genuine+malicious (label-flipping) ($T_{th}$ = $0.5$) & 0.9896 & 0.9721  & 0.9933 & 0.9826  \\\cline{3-7}
    & & w/genuine+malicious (policy induction) ($T_{th}$ = $0.5$) & 0.9918 & 0.9764 & 0.9963  & 0.9862  \\\cline{3-7}
    & & w/genuine only ($T_{th}$ = $0.8$) & 0.9908  & 0.9730  & 0.9966 & 0.9847  \\\cline{3-7}
    & & Baseline (no transfer) & 0.9474 & 0.8944  & 0.9317  & 0.9127  \\\cline{3-7}
    \hhline{=======}
    \multirow{8}{*}{SC2} & \multirow{4}{*}{DoS Random Sybil} & w/genuine+malicious (label-flipping) ($T_{th}$ = $0.5$) & 0.5659  & 0.5616 & 0.9915 & 0.7144  \\\cline{3-7}
    & & w/genuine+malicious (policy induction) ($T_{th}$ = $0.5$) & 0.5690 & 0.5606  & 0.9873  & 0.7150  \\\cline{3-7}
    & & w/genuine only ($T_{th}$ = $0.8$) & 0.5656 & 0.5581 & 0.9929 & 0.7145  \\\cline{3-7}
    & & Baseline (no transfer) & 0.5078  & 0.4730 & 0.5108  & 0.4912  \\\cline{3-7}
    \hhline{|~======}
    & \multirow{4}{*}{DoS Disruptive Sybil} & w/genuine+malicious (label-flipping) ($T_{th}$ = $0.5$) & 0.5544  & 0.5552  & 0.9955  & 0.7129  \\\cline{3-7}
    & & w/genuine+malicious (policy induction) ($T_{th}$ = $0.5$) & 0.5523  & 0.5544  &  0.9898  & 0.7107   \\\cline{3-7}
    & & w/genuine ($T_{th}$ = $0.8$)  & 0.5552  & 0.5557  & 0.9960  & 0.7134  \\\cline{3-7}
    & & Baseline (no transfer) & 0.4928 & 0.4587 & 0.5018  & 0.4793  \\\cline{3-7}
    \hhline{=======}
    \multirow{8}{*}{SC3} & \multirow{4}{*}{Position variants*} & w/genuine+malicious (label-flipping) ($T_{th}$ = $0.5$) & 0.9224 & 0.7982 & 0.9856  & 0.8822  \\\cline{3-7}
    & & w/genuine+malicious (policy induction) ($T_{th}$ = $0.5$) & 0.9195 & 0.7956 & 0.9779 & 0.8774  \\\cline{3-7}
    & & w/genuine only ($T_{th}$ = $0.8$) & 0.9236  & 0.7998  & 0.9880  & 0.8840  \\\cline{3-7}
    & & Baseline (no transfer) &  0.4978 & 0.2956  & 0.5095  & 0.3741  \\\cline{3-7}
    \hhline{|~======}
    & \multirow{4}{*}{Speed variants**} & w/genuine+malicious (label-flipping) ($T_{th}$ = $0.5$) & 0.9447 & 0.8540  & 0.9824 & 0.9135  \\\cline{3-7}
    & & w/genuine+malicious (policy induction) ($T_{th}$ = $0.5$) & 0.9348 & 0.8275  & 0.9857 & 0.8997  \\\cline{3-7}
    & & w/genuine only ($T_{th}$ = $0.8$) & 0.9394 & 0.8399 & 0.9845  & 0.9062  \\\cline{3-7}
    & & Baseline (no transfer) & 0.4988  & 0.2948  &  0.4953 & 0.3696  \\\cline{3-7}
    \hline
    \multicolumn{2}{l}{*\footnotesize{Constant Position Offset, Random Position, and Random Position Offset}} &  
    \multicolumn{2}{l}{**\footnotesize{Constant Speed Offset, Random Speed, and Random Speed Offset}} \\
\end{tabular}}
\end{table*}

Detection performance in Table~\ref{tab:misdet_perf_results} reveals 
that target learning with knowledge transfer yields significantly superior F-scores compared to the baseline scheme in SC3. Under both $T_{th}$ values, the F-scores of $0.88$ and slightly over $0.90$ for position- and speed-related misbehaviors, respectively, demonstrate high effectiveness in detecting a partially observable attack space. Conversely, the baseline scheme %
becomes highly ineffective, as shown by the low F-score of $0.37$ with high number of FPs and FNs for both position- and speed-related misbehaviors. It is worth \CHANGED{noting} %
that misbehavior variants generated by adding/subtracting an offset (i.e., Constant/Random Position Offset and Constant/Random Speed Offset) are more challenging to detect as compared to others, such as Constant Position and Constant Speed. Thus, the transfer of relevant knowledge to the target RSU \CHANGED{becomes imperative} to effectively identify such partially observable attack spaces. %

\begin{figure}[!t]
\centering
\includegraphics[width=0.48\textwidth]{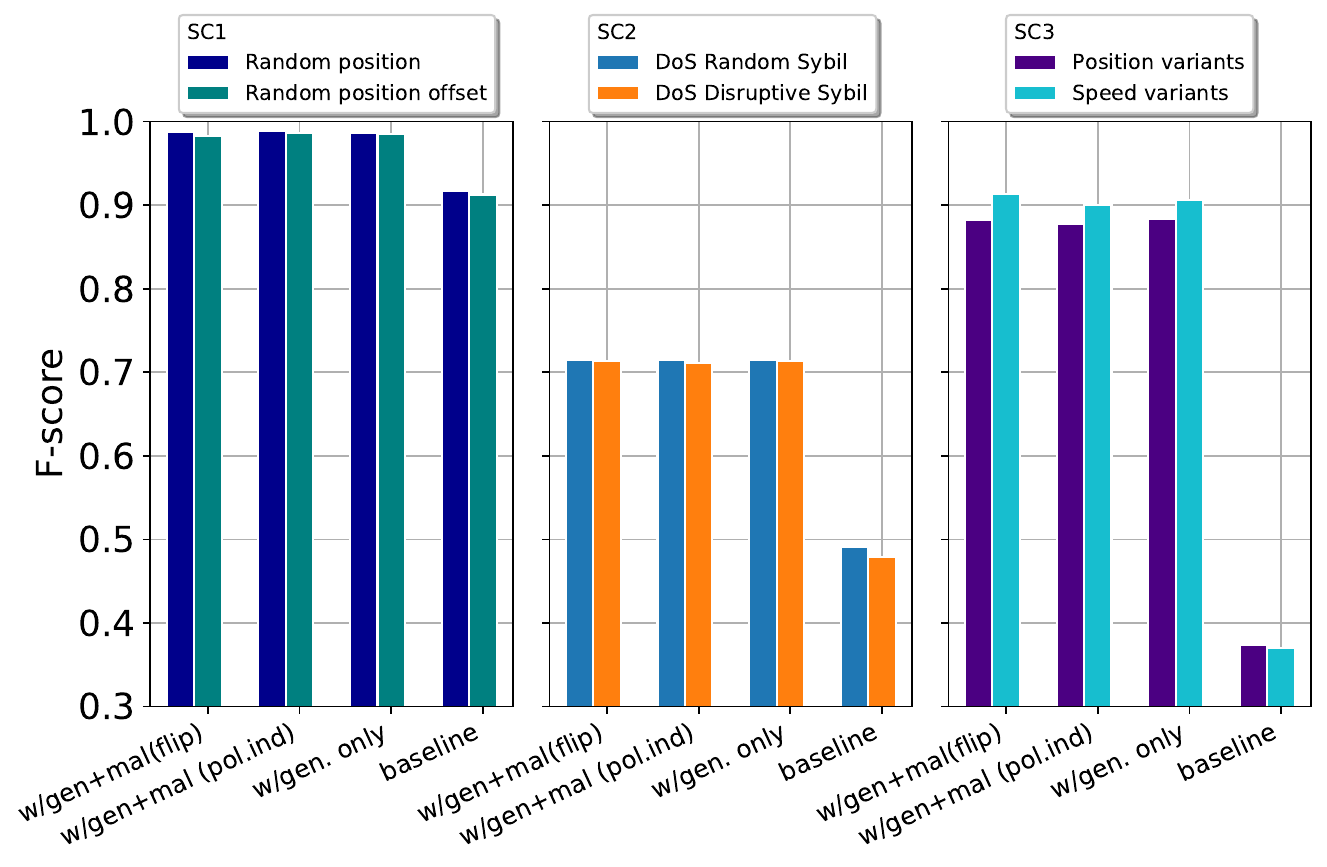}
\caption{Impact of knowledge transfer based on F-scores. Transfers feature both genuine and malicious \CHANGED{(i.e., label-flipping and policy induction attacks)} sources when $T_{th}$ = $0.5$, %
only genuine sources involve when $T_{th}$ = $0.8$
}
\label{fig:f1_score_comparison}
\vspace*{0.1cm}
\end{figure}

Overall, across \CHANGED{all} three scenarios, collaborative misbehavior detection with knowledge transfer significantly outperforms 
tabula rasa learning, as \CHANGED{summarized} %
in Fig.~\ref{fig:f1_score_comparison}, demonstrating its high effectiveness. Specifically, our approach enhances robustness and generalizability by effectively detecting previously unseen and partially observable misbehavior attacks.

\subsection{Discussion}

\REVISED{In summary, we develop a selective knowledge transfer approach that selects knowledge learned at source RSUs with high semantic relatedness to the target RSU, enabling effective and efficient learning about various misbehaviors in a collaborative fashion. We introduce a relevant theoretical framework for the proposed selective knowledge transfer in DRL-based collaborative misbehavior detection in untrusted vehicular environments. In addition, empirical evidence demonstrating the necessity of transfer learning is provided across three examined scenarios discussed in Table~\ref{tab:attack_scenarios} through $(i)$ asymptotic improvement in the learning performance of the target RSU (Figs.~\ref{fig:rndpos_rndposoffset_target_training}--\ref{fig:sc3_position_speed_target_training}); $(ii)$ improved misbehavior detection performance at the target RSU, as shown in Table~\ref{tab:misdet_perf_results}; and $(iii)$ faster learning speed of the target RSU, as shown in Table~\ref{tab:training_time_reduction}.}

\REVISED{More specifically}, we empirically demonstrate that the selective knowledge transfer approach, coupled with trust evaluation, significantly reduces the training time for collaborative misbehavior detection at the target RSU and produces key outcomes: $i)$ higher performance levels in SC1 and SC3 for the detection of future attacks of the same type and only partially observable misbehaviors, respectively, and $ii)$ a certain performance level in SC2 by effectively detecting unseen attacks. 
\CHANGED{Our} experiments further demonstrate that the proposed knowledge transfer \CHANGED{method} %
is robust across various threshold values ($T_{th} \in [0,1]$) when \CHANGED{selecting} %
source RSUs. \CHANGED{In particular,} Figs.~\ref{fig:rndpos_rndposoffset_target_training}--%
\ref{fig:f1_score_comparison} consistently demonstrate similar learning and detection performances, irrespective of specific $T_{th}$ values used. Nonetheless, it is worth mentioning that appropriate threshold selection depends on the severity of the adversarial attack model and the specific scenario being examined. 

The %
generalizability problem is an essential challenge in DL- and \CHANGED{traditional} ML-based misbehavior detection methods proposed in related work (Sec.~\ref{sec:related_work}). \CHANGED{Conventional} approaches often struggle with detecting unseen or only partially observable attack spaces. In contrast, our collaborative misbehavior detection approach leverages transfer learning to improve generalizability by effectively handling both unseen and partially observable attacks.
A concern \CHANGED{unavoidably} arises with the increased computational overhead in the form of processing and buffer sizes at the target \CHANGED{RSU} when resorting to more tolerant $T_{th}$ values. Lower threshold values result in the inclusion of more selected sources, which may encompass less trustworthy ones. Consequently, this poisons the experience samples buffer $\tilde{\mathcal{S}}$ with samples from sources with lower trust values, incurring processing overhead on the target \CHANGED{RSU}. %
However, samples from less trustworthy sources will not \CHANGED{eventually} influence the target's learning process due to the experience selection.

\section{Conclusions}
\label{sec:conclusion}
The current research focus in AI/ML-based collaborative misbehavior detection lies in designing methods that can effectively identify various misbehavior attacks. \CHANGED{Such approaches, albeit resonating high detection accuracy, assume the presence of an honest majority, with inherent model resilience and without explicit defense against} %
adversarial attacks. %
While the majority of solutions primarily concentrate on centralized setups, they often suffer from drawbacks related to latency, computational cost, scalability, robustness, and generalizability, due to the heterogeneity and distributed nature of vehicular environments.  
To address these research challenges, we proposed a DRL-based scheme %
that utilizes transfer learning for distributed collaborative misbehavior detection \CHANGED{among RSUs}. Our approach enables selective knowledge transfer from reliable source \CHANGED{RSUs} in unpredictable and untrusted vehicular environments by leveraging semantic relatedness with the target RSU.
We empirically showcase that selective knowledge transfer, coupled with trust evaluation, is sample-efficient and effective in detecting previously unseen and partially observable misbehaviors with high detection performance. %
Hence, significant steps towards generalizability are attained.

In future research, we intend to delve into adversarial training techniques for DRL\CHANGED{, such as introducing perturbations to the observation space,} to counter gradient-based adversarial attacks within target RSUs. Moreover, we are interested in evaluating the scalability and performance of our approach when compared to emerging blockchain-based trust management \CHANGED{techniques} in vehicular networks.

\vspace*{-0.3cm}
{\appendix[Misbehavior Attack Types]\label{sec:apx_attacks}

\noindent\textbf{Position falsification:} A vehicle transmits falsified position coordinates, leading to four different %
variants: i) \textit{constant position}, the attacker transmits fixed position coordinates; ii) \textit{constant position offset}, the attacker transmits the real position coordinates with a fixed offset; iii) \textit{random position}, the attacker transmits newly generated random position coordinates; and iv) \textit{random position offset}, the attacker transmits the real position coordinates with a random offset.

\vspace{0.02in}
\noindent\textbf{Speed falsification:} A vehicle transmits falsified speed values in its BSM, following a similar approach as in position falsification attack. This results in \textit{constant speed, constant speed offset, random speed} and \textit{random speed offset} %
variants.

\vspace{0.02in}
\noindent\textbf{Disruptive:} This is similar to a data replay attack, where a vehicle re-transmits previously sent messages by other vehicles. BSMs are selected at random and flood the network with stale data to disrupt genuine information from being propagated. This attack may also be carried out in \textit{DoS} and \textit{Sybil} modes. 

\vspace{0.02in}
\noindent\textbf{Denial-of-service (DoS):} A misbehaving vehicle transmits BSMs at a very higher frequency than the acceptable limit set by the standard. This results in high volume of data transmission causing extensive periods of network congestion and unavailability of critical services to legitimate vehicles. 

\vspace{0.02in}
\noindent\textbf{DoS Random:} In this attack, the attacker sets all BSM fields to random values and performs a typical DoS attack. 

\vspace{0.02in}
\noindent\textbf{DoS Disruptive:} An attacker may re-transmit previously sent BSMs by other legitimate vehicles. BSMs are selected at random and flood the network with stale data with the intention of disrupting genuine information from being propagated. The attacker increases BSM transmission rate to realize this attack.

\vspace{0.02in}
\noindent\textbf{DoS Random Sybil:} The attacker changes pseudonym identities on every %
BSM while performing the DoS random attack. 

\vspace{0.02in}
\noindent\textbf{DoS Disruptive Sybil:} The attacker constantly changes pseudonyms on every re-transmission of BSMs while concealing its real identity.

\bibliographystyle{IEEEtran}

\bibliography{main}

\end{document}